\documentclass[aps,pre,preprint,floatfix,notitlepage]{revtex4-1}
\newcommand{\be}{\begin{eqnarray}}
\newcommand{\ee}{\end{eqnarray}}
\usepackage{graphicx}
\usepackage{verbatim}
\usepackage{amsmath}
\usepackage{enumerate}
\usepackage{subcaption}

\begin{document}
\title{A numerical study of longtime dynamics and ergodic-nonergodic transitions in dense simple fluids}
\author{David D. McCowan}
\affiliation{The James Franck Institute and the Department of Physics, The University of Chicago, Chicago, Illinois 60637, USA}
\date{\today}

\begin{abstract}
For over 30 years, mode-coupling theory (MCT) has been the \textit{de facto} theoretic description of dense fluids and the transition from the fluid to glassy state. MCT, however, is limited by its \textit{ad hoc} construction and lacks a mechanism to institute corrections. We use recent results from a new theoretical framework---developed from first principles via a self-consistent perturbation expansion in terms of an effective two-body potential---to numerically explore the kinetics of systems of classical particles, specifically hard spheres governed by Smoluchowski dynamics. We present here a full solution for such a system to the kinetic equation governing the density-density time correlation function and show that the function exhibits the characteristic two-step decay of supercooled fluids and an ergodic-nonergodic transition to a dynamically-arrested state. Unlike many previous numerical studies,---and in stark contrast to experiment,---we have access to the full time and wavenumber range of the correlation function with great precision, and are able to track the solution unprecedentedly close to the transition, covering nearly 15 decades in scaled time. Using asymptotic approximation techniques analogous to those developed for MCT, we fit the solution to predicted forms and extract critical parameters. Our solution shows a transition at packing fraction $\eta^* = 0.60149761(10)$---consistent with previous static solutions under this theory and with comparable colloidal suspension experiments---and the behavior in the $\beta$-relaxation regime is fit to power-law decays of the typical forms with critical exponents $a = 0.375(3)$ and $b=0.8887(4)$, and critical exponent parameter $\lambda = 0.5587(18)$. For the $\alpha$-relaxation of the ergodic phase, we find a power-law divergence of the time scale $\tau_{\alpha}$ as we approach the transition, and in the nonergodic phase we find glass form factors whose amplitudes scale as the square-root of distance from the transition as predicted. Through these results, we establish that this new theory is able to reproduce the salient features of MCT, but has the advantages of being derived from first principles and possessing a clear mechanism for making systematic improvements.
\end{abstract}

\maketitle

\section{Introduction}
\label{sec:introduction}
It is possible to compress or cool a fluid beyond the freezing point where it would ordinarily crystalize into a solid and instead form a supercooled liquid or, eventually, a glass. As one approaches this glassy state, the dense fluid can show several features unique to this transition: the relaxation dynamics slow, the structure arrests, and a disordered non-equilibrium state emerges\cite{stillinger,ediger,angell00,goetze92}. This glass transition is seen in a wide variety of fluids from complex molecular liquids\cite{dixon, borjesson, petry, li1, li2, sidebottom, halalay, debenedetti, chong} to simple colloidal suspensions\cite{pusey87,vanmegen98,cheng,phan}.

When studying liquids and the transition from liquid to glass, the natural quantity to monitor is the the density-density time correlation function,
\be
G_{\rho\rho}(q,t) = \langle\delta\rho({\bf q},t)\delta\rho(-{\bf q},0)\rangle,
\ee
where ${\bf q}$ is wavenumber, $t$ is time, $\delta\rho({\bf q},t)$ is the particle density fluctuation, and the angle brackets represent the canonical ensemble average. This function---also known as the intermediate static structure factor or (in the context of light or neutron scattering experiments), the intermediate scattering function---captures the relaxation of the system away from the static structure factor $S(q)$ which characterizes the equilibrium static state,
\be
S(q) = G_{\rho\rho}(q,t=0).
\ee

\begin{figure}[p]
\includegraphics[width=\columnwidth]{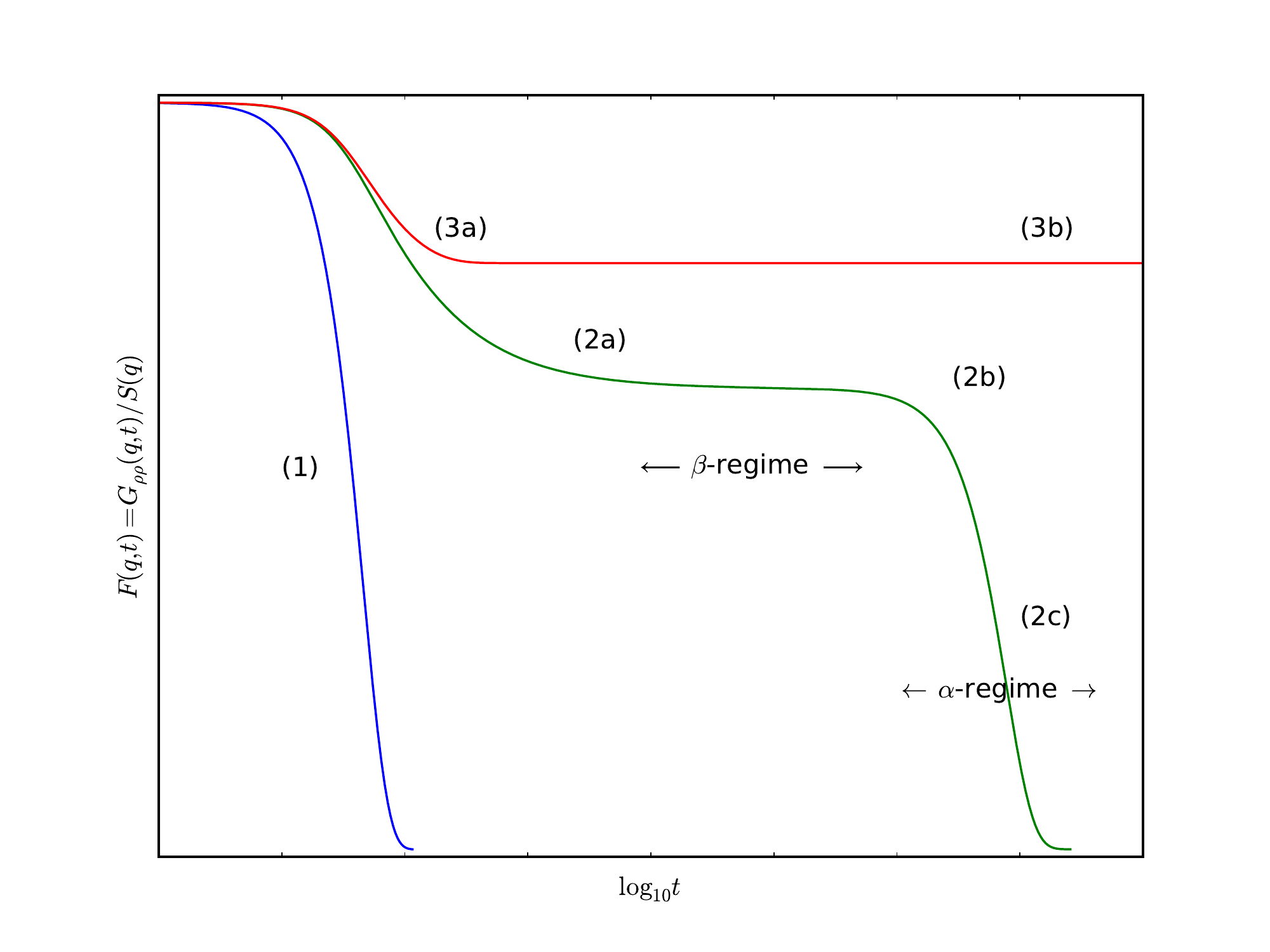}
\caption{The general features of the intermediate structure factor are shown for (1) an ergodic system far from the transition, (2) an ergodic system close to the transition, and (3) and a nonergodic system. Far from the transition, the ergodic liquid shows pure exponential decay, but as it approaches the transition, the characteristic two-step decay emerges. First, the system settles into the $\beta$-relaxation regime with a power-law decay into the plateau proportional to $t^{-a}$ (2a) and a power-law decay out of the plateau proportional to $t^b$ (2b). This latter decay is termed the von Schweidler relaxation and marks both the end of the $\beta$-relaxation and the beginning of the $\alpha$-relaxation regime where the system enters its final (possibly stretched) exponential decay (2c). In the nonergodic regime, the system relaxes by the same $t^{-a}$ power law into the plateau (3a) and remains frozen at that value (3b) even at long times.}
\label{fig:schematic}
\end{figure}

In a dilute liquid, this function decays exponentially with time, but as one approaches the liquid-glass transition, there is a significant slowing down and the density-density time correlation function evolves to show non-Arhenius, stretched exponential behavior. Closer to the transition, this evolves further into a two-step decay; at intermediate times the system remains nearly stationary, forming a plateau which can extend decades in time before finally relaxing to zero. Beyond the transition point, the system becomes nonergodic and the ability of particles to diffuse through the whole of the system is halted. The general features of the intermediate structure factor are schematically outlined in Fig. \ref{fig:schematic}.

We may identify two relaxation regions of interest. First, the relaxation into and out of the extended plateau is called the $\beta$-relaxation and is due to the caging effect; though free to move on very short time scales, particles are locally confined to ``cages" by neighboring particles and the time scale for escaping such a cage, $\tau_{\beta}$, grows with increasing density. Next, the eventual relaxation out of the plateau and to zero is called the $\alpha$-relaxation; the time scale associated with this (possibly stretched) exponential decay is $\tau_{\alpha}$ and also grows with increasing density. Both these time scales diverge at the transition, leaving the system in the nonergodic state and the intermediate structure factor unable to relax beyond the the plateau.

Though the glassy state is a solid, it is amorphous and lacks the long range order of the equilibrium crystalline state. This dynamic arrest manifests itself in the longtime limit of the intermediate structure factor as a discontinuous jump from zero to the fixed plateau amplitude. The positive, nonzero limit of the normalized correlator $F(q,t) = G_{\rho\rho}(q,t)/S(q)$ is called the glass form factor, the plateau, or the nonergodicity factor,
\be
\lim_{t\rightarrow\infty}F(q,t) = f_q \geq 0.
\ee
The nonergodicity factor also plays the role of the so-called Debye-Waller factor as it appears as the amplitude of the zero wavenumber peak in the real part of the Fourier-transformed spectrum in the glassy phase,
\be
F^{\prime\prime}(q,\omega)=\pi f_q\delta(\omega) + \textrm{regular terms}.
\ee
In the ergodic liquid phase, this $\delta$-function peak broadens to a quasi-elastic peak and
the area of such a peak can be considered an effective Debye-Waller factor\cite{goetze99}.

Though we may speak theoretically of longtime limits and ergodic-nonergodic transitions, the situation is less clear cut in experiment and simulation. There, the glass transition must be defined in an operational sense, with the most common definition of the calorimetric glass transition temperature $T_g$ (or density $\eta_g$) as the temperature (or density) at which the viscosity increases beyond $10^{13}$ Poise\cite{angell84}; this is a somewhat arbitrary choice, but one which represents an increase in the relaxation time of the system so substantial that it can no longer reasonably be measured in the lab.

Despite the fuzziness of the definition, the calorimetric glass transition is in practice rather sharp and we find that $T_g$ forms a robust scale characterizing the system. Remarkably, many dynamic quantities which appear to diverge at the glass transition---e.g., viscosity, diffusion constant, and relaxation time---can be collapsed to universal functions when temperature is scaled by $T_g$ (or density scaled by $\eta_g$)\cite{angell84}. Likewise, the two-step decay of the density-density correlation function detailed above appears on supercooling for so wide a range of materials---from macroscopic colloidal suspensions described by Brownian dynamics\cite{pusey87,vanmegen98,cheng,phan} to complex molecular or polymer fluids whose behavior is controlled by atomic-level forces\cite{dixon, borjesson, petry, li1, li2, sidebottom, halalay, debenedetti, chong}---that it is considered the very hallmark of the glass transition.

The first successful model of the glass transition was mode-coupling theory (MCT) introduced in the 1980's by Leutheusser and Goetze\cite{leutheusser,goetze84,bengtzelius,goetze85}. MCT begins with exact equations of motion, then uses a projection operator formalism\cite{mori} to integrate out the ``fast" variables, leaving behind only the ``slow" variables responsible for the glassy dynamics. Such a projection leads to a generalized Langevin equation (or kinetic equation) for generic correlation function $C(t)$ of the form
\be
\frac{\partial^2 C(q,t)}{\partial t^2} = -\Omega^2(q) C(q,t) - \int_0^t ds K(t-s)\frac{\partial C(q,s)}{\partial s}
\ee
where $K(q,t)$ is the so-called memory function and $\Omega^2(q)$ is a time-independent quantity that depends on the nature of the correlation function $C(q,t)$\cite{andreanov,goetze90}.

This equation, however, is just a rearrangement and is equally as complex as the original problem. The key development of MCT was to move to rederive this in terms of a set of fluctuating variables and then approximate the memory function by factorizing four-point correlation functions into products of two-point correlation functions. In the case of the density-density time correlation function, this came to be known as the mode-coupling approximation\cite{goetze92} and the memory function takes the form
\be
K(q,t) = \frac{1}{2}\int \frac{d^d {\bf k}}{(2\pi)^d} M^2({\bf q},{\bf k})G_{\rho\rho}(k,t)G_{\rho\rho}(|{\bf q}-{\bf k}|,t)
\label{eq:MCT}
\ee
where
\be
M^2({\bf q},{\bf k}) = \frac{\rho}{q^2}\bigg[{\bf \hat{q}}\cdot{\bf k} V(k)
+ {\bf \hat{q}}\cdot|{\bf q} - {\bf k}|V(|{\bf q} - {\bf k}|)\bigg]^2
\ee
is the vertex function, $V(q)$ is the pair-wise potential, and $\rho$ is the average particle density. This approximation is not justified in any rigorous sense, but solutions to the kinetic equation nonetheless show the two-step decay for dense fluids and predict an ergodic-nonergodic transition at finite density.

Mode-coupling theory has undeniably had a number of successes. In addition to giving the correct qualitative form for the two-step decay, MCT makes remarkable predictions about power-law relaxation into and out of the plateau in the $\beta$-regime as well as diverging time-scales that have held up to experimental verification. The MCT predictions for the nonergodicity factor $f_q$ have also matched measured quantities\cite{goetze92, goetzerev, goetze99, das, goetzebook}.

MCT, however, has also has had a number of shortcomings. Most pointedly, MCT does not accurately predict transition temperatures and densities, is not derived from first principles, and has no clear method for controlled improvements\cite{reichman,andreanov}.

For this reason, there is great interest in a new theory which improves over MCT. There have been a number of previous attempts\footnote{Initial efforts to ``derive” MCT using kinetic theory were only crudely successful. In order to gain some control, effective dynamical field theories were introducted, begining with the realization that MCT can be understood in terms of the fluctuating nonlinear hydrodynamics of dense fluids\cite{toner, dasmazenko, dasmazenko09}. In this approach, however, the formal structure of the field theory is sufficiently complex that it is difficult to go beyond one-loop order in the calculation. Next came the Dean-Kawasaki model\cite{kaw95,dean,kawasaki,miyazaki,kimkawasaki}, the simplest field theoretic model that describes the kinetics of the colloidal systems operating under Smoluchowski dynamics. It too, however, ran aground as it is difficult to establish whether the model supports an ergodic-nonergodic transition even at one loop order. Other, more complicated field theoretic models have also been put forth\cite{mazenkoarxiv,RDM1,RDM2,ABL,cates,cristani,szamelFT1,szamelFT2}. The use of effective field theories turns out to be technically as complicated as a microscopic approach with the draw back of not reproducing the correct statics and the need to introduce a large wavenumber cutoff. What was needed was the first principles fully self-consistent formulation of kinetic theory seen in the theory of this paper.} which have fallen short, but we here expand on a promising recent theory\cite{FTSPD, SDENE, NDI, NDII, ULTD} and use it to develop the full dynamic results in a model system of hard spheres obeying Smoluchowski dynamics.

The theory---a field theoretic approach that derives the intermediate structure factor (and other associated correlation functions) in a self-consistent perturbation expansion in the pair-wise potential---yields a kinetic equation and memory function very similar to that of MCT at second order in the potential expansion\cite{KEGSD} and produces all the expected features near the ergodic-nonergodic transition. The full numerical solution shows a two-step relaxation, diverging length scales $\tau_{\alpha}$ and $\tau_{\beta}$, power-law decay into and out of the plateau in the $\beta$-regime, and scaling of the amplitudes of the nonergodicity factor, $f_q$\cite{spyridis}. The value we find for the transition density is in rough agreement with relevant experiment and simulation (unlike MCT's)\cite{brambilla}, and our power-law decays obey the predicted forms, though with parameter values slightly different than those seen elsewhere for hard spheres\cite{goetze99}. Likewise, our nonergodicity factors $f_q$ show similarity to, but not quantitative agreement with measured results\cite{goetze99}.

The structure of this paper is as follows.

In Section \ref{sec:theory}, we introduce the theory and motivate the governing equations. While the connection to the work of mode-coupling theory occurs at second order in the potential expansion, we first briefly review results at zeroth and first orders in order to show how new physics are introduced order-by-order. In addition to the governing equations, we also outline the asymptotic analysis which makes predictions for the critical dynamics near the transition.

In Section \ref{sec:solution}, we define the system we study---hard spheres obeying Smoluchowski dynamics---and walk through a numerical solution to the kinetic equation. We highlight features of the intermediate structure factor at different packing fraction densities and wavenumber and find the two-step decay in the ergodic phase giving way to a clear transition to the nonergodic phase. Through fits of the data, we explore the divergence of the $\alpha$- and $\beta$-relaxation times and extract values for the critical power-law exponents.

Finally, we compare these results to mode-coupling theory, experiment and simulation in Section \ref{sec:comparison}.

Unlike many previous numerical studies,---and in stark contrast to experiment,---this study yields the full time and wavenumber range of the correlation function with great precision. As such, we are able to track the solution unprecedentedly close to the transition, covering nearly 15 decades of scaled time and having access to very detailed data against which to fit and test predicted forms.

It is worth stressing that this work represents the first numerical solution to display the full dynamics near an ergodic-nonergodic transition derived from a theory outside mode-coupling theory. As the product of a well-motivated and self-consistent perturbation expansion, this theory reproduces many of the characteristic features of the glass-transition, and places it in a position to supersede MCT as the tool of choice for studying supercooled fluids.

\section{Theory}
\label{sec:theory}
A new fundamental theory of statistical particle dynamics that unifies kinetic theory, Brownian motion, and field theory techniques was developed in Ref. \onlinecite{FTSPD} for Smoluchowski dynamics and extended in Ref. \onlinecite{NDI} to Newtonian dynamics. Here, we review the main points.

\subsection{Fundamental theory of statistical particle dynamics}
Imagine a system of $N$ particles of mass $m$ located at positions $R_i(t)$ and possessing momenta $P_i(t)$. Such particles interact via a pair potential
\be
U = \frac{1}{2}\sum_{i\neq j} V(R_i-R_j)
\label{eq:potential}
\ee
which leads to a force
\be
F_i =-\nabla_i U.
\ee
Under Newtonian dynamics, the equations of motion are
\be
m\dot{R}_i = P_i
\ee
and
\be
\dot{P}_i = F_i.
\ee
If, however, the particles interact with a thermal bath such that the individual momenta are quickly dissipated (making individual particle acceleration negligible), the system can instead by described by simpler Smoluchowski dynamics. In this case, we have only the Langevin equation given by
\be
\dot{R}_i = DF_i + \eta_i
\ee
where $D$ is the diffusion coefficient and $\eta_i(t)$ is Gaussian-distributed random noise with zero mean
\be
\langle \eta_i(t) \rangle = 0
\ee
and variance proportional to temperature
\be
\langle \eta_i(t)\eta_j(t^{\prime}) \rangle = 2D\beta^{-1}\delta_{ij}\delta(t-t^{\prime})
\ee
where $\beta^{-1} = k_BT$ and where the angle brackets here represent canonical averages\footnote{As was shown in Ref. \onlinecite{ULTD}, the theory predicts that the short-time dynamics of these two types of systems will be quite different (ballistic versus Brownian), but converge to the same long-time dynamics. This is consistent with simulations\cite{gleim,szamel04,hofling}. Importantly for our development here, the differences between the two dynamics types can all be rolled into the definition of the response field and the calculation of the zeroth-order cumulants and vertices; the remaining formalism is identical.}.

We wish to move from these equations of motion to a field theory. One can form a Martin-Siggia-Rose (MSR) action\cite{msr1,msr2} which leads to a grand canonical partition function. For a set of core dynamical fields $\Phi = \{\Phi_{\alpha}\}$, we have
\be
Z_T = \sum_{N=0}^{\infty}\frac{\rho_0^N}{N!}\textrm{Tr}e^{-A_I + H\cdot\Phi}.
\ee
where $\rho_0$ is the fugacity (or bare density) and where the interacting terms of the action are given by
\be
A_I = \frac{1}{2}\sum_{\alpha\beta}\int d{\bf x}_1 dt_1 d{\bf x}_2 dt_2
\Phi_{\alpha}({\bf x}_1,t_1)\sigma_{\alpha\beta}({\bf x}_1,{\bf x}_2,t_1,t_2)\Phi_{\beta}({\bf x}_2,t_2)
\ee
or, in shorthand,
\be
A_I = \frac{1}{2}\Phi_{\alpha}(1)\sigma_{\alpha\beta}(12)\Phi_{\beta}(2),
\ee
such that repeated subscripts are summed over and repeated space and time arguments are integrated over. The non-interacting terms of the action, $A_0 = A - A_I$, and the initial conditions have been rolled into the trace and we allow for coupling to external fields $H=\{H_{\alpha}\}$ via a Zeeman-like term,
\be
H_{\alpha}(1)\Phi_{\alpha}(1) = \sum_{\alpha}\int d{\bf x}_1 dt_1 H_{\alpha}({\bf x}_1,t_1)\Phi_{\alpha}({\bf x}_1,t_1).
\ee
Details of the development of the action and the trace are discussed more carefully in Ref \onlinecite{FTSPD}.

There are two essential physical fields which appear in the Hamiltonian and which therefore define what we will term the \textit{core} problem. These two fields -- the density $\rho({\bf x},t)$ and the response field $B({\bf x},t)$ -- must always be included,
\be
\Phi(1) = \{\rho({\bf x_1},t_1), B({\bf x_1}, t_1),\ldots\},
\ee
and couple to each other via the interaction matrix,
\be
\sigma_{\alpha\beta}(12) = V({\bf x_1}-{\bf x_2})\delta(t_1-t_2)
[\delta_{\alpha\rho}\delta_{\beta B} + \delta_{\alpha B}\delta_{\beta \rho}]
\ee
where $V({\bf x_1}-{\bf x_2})$ is the same pair potential defined above in Eq.~(\ref{eq:potential}). While these two fields cover essentially all the degrees of freedom of interest in a Smoluchowski system, the momentum degrees of freedom of a Newtonian system lead to additional conservation laws and one may want to look in such a system, for example, at couplings to momentum current (either the whole current or the two transverse components), kinetic energy density, or phase space density. We will not discuss these extensions here, though an introduction is given in Ref. \onlinecite{NDI}.

The density $\rho({\bf x},t)$ is of the usual form for both types of dynamics,
\be
\rho({\bf x},t) = \sum_{i=1}^N \delta({\bf x}-{\bf R}_i(t)),
\ee
however the response field $B({\bf x},t)$ which emerges is unique to this theory. In the Newtonian dynamics case it takes the form
\be
B^{(ND)}({\bf x},t) = \sum_{i=1}^N \hat{P}_i(t)i\nabla\delta[{\bf x}-{\bf R}_i(t)]
\ee
where $\hat{P}_i(t)$ is the MSR conjugate momentum field, and in the Smoluchowski dynamics case it takes the form
\be
B^{(SD)}({\bf x},t) = D\sum_{i=1}^N [\hat{R}_i(t)\nabla +\theta(0)\nabla^2]\delta[{\bf x}-{\bf R}_i(t)]
\ee
where $\hat{R}_i(t)$ is the MSR conjugate position field. The response field $B$ is key to the development of this theory and marks an important break with MSR tradition where the conjugate position and momenta fields themselves play a response role.

From the partition function, we may construct the generating functional
\be
W[H] = \ln Z_T[H]
\ee
and form cumulants by taking successive functional derivatives with respect to the coupling field,
\be
G_{\alpha}({\bf x_1},t_1)
&=& \langle \delta\Phi_{\alpha} \rangle
= \frac{\delta}{\delta H_{\alpha}} W[H],
\nonumber\\
G_{\alpha\beta}({\bf x_1},t_1;{\bf x_2},t_2)
&=& \langle \delta\Phi_{\alpha} \delta\Phi_{\beta} \rangle
= \frac{\delta}{\delta H_{\beta}} G_{\alpha}
= \frac{\delta}{\delta H_{\alpha}}\frac{\delta}{\delta H_{\beta}}W[H]
\nonumber\\
G_{\alpha\beta\gamma}({\bf x_1},t_1;{\bf x_2},t_2; {\bf x_3},t_3)
&=& \langle \delta\Phi_{\alpha} \delta\Phi_{\beta} \delta\Phi_{\gamma}\rangle
= \frac{\delta}{\delta H_{\gamma}} G_{\alpha\beta}
= \frac{\delta}{\delta H_{\alpha}}\frac{\delta}{\delta H_{\beta}}\frac{\delta}{\delta H_{\gamma}} W[H]
\nonumber\\
&&\ldots
\ee
Keeping the coupling fields can, for example, allow one to treat trapped or driven systems, but we will here only be concerned with the steady-state cumulants where all $H_{\alpha}$ are set to zero after taking the derivatives.

Though the above is exact, it is a formal development. In order to make traction, one must perform a series expansion to compute forms for the cumulants order-by-order. The key identity\cite{FTSPD} is
\be
G_{\alpha} = \textrm{Tr}\phi_{\alpha} e^{H\cdot\phi+\Delta W[H]}
\ee
where
\be
\Delta W[H] &=& \phi_{\alpha}(1)\sigma_{\alpha\beta}(12)G_{\beta}(2)
\nonumber\\
&&+ \frac{1}{2}\phi_{\alpha}(1)\phi_{\beta}(2)\sigma_{\alpha\gamma}(13)\sigma_{\beta\delta}(24)G_{\gamma\delta}(34)
\nonumber\\
&&+ \frac{1}{3!}\phi_{\alpha}(1)\phi_{\beta}(2)\phi_{\gamma}(3)
\sigma_{\alpha\delta}(14)\sigma_{\beta\epsilon}(25)\sigma_{\gamma\zeta}(36)G_{\delta\epsilon\zeta}(456)
\nonumber\\
&&+\ldots
\ee
and where
\be
\Phi_{\alpha}({\bf x},t)=\sum_{i=1}^N \phi_{i}({\bf x},t).
\ee
Each derivative brings down a factor of $\Delta W[H]$ which can be truncated to any order in the interaction matrix $\sigma_{\alpha\beta}(12) \sim V({\bf x}_1-{\bf x}_2)$. In this way, we have now cast the problem as a self-consistent perturbation expansion in the potential. Once one solves for the zeroth-order (i.e. non-interacting) cumulants, -- a non-trivial endeavor, -- higher orders are generated by turning the crank.

\subsection{Zeroth-order cumulants and vertex functions}
The zeroth-order cumulants differ for Newtonian and Smoluchowski dynamics, but the formalism of extending order-by-order is identical.

In the case of Smoluchowski dynamics, the zeroth order cumulants can be derived as
\be
G^{(0)}_{B\ldots B \rho\ldots\rho}(1,\ldots, \ell, \ell+1, \ldots, n)=
\rho_{0}(2\pi )^{d}\delta \bigg(\sum_{i=1}^{n}{\bf q}_i\bigg)b(1)\ldots b(\ell)
e^{N_{n}}
\label{eq:cumulant}
\ee
where
\be
N_{n}=\frac{1}{2}\bar{D}\sum_{i=1}^{n}\sum_{j=1}^{n}{\bf q}_i\cdot {\bf q}_j|t_{i}-t_{j}|
\ee
and
\be
b(j)=D\sum_{i\neq j=1}^{n} {\bf q}_i\cdot {\bf q}_j\theta (t_{i}-t_{j}),
\ee
and where $\bar{D}=\beta^{-1}D$. Explicitly, the two-point cumulants are
\be
G^{(0)}_{\rho\rho}(12)&=&\rho_0(2\pi)^d\delta({\bf q}_1+{\bf q}_2)e^{-\bar{D}q_1^2|t_1-t_2|},
\\
G^{(0)}_{\rho B}(12)&=&-\rho_0(2\pi)^d\delta({\bf q}_1+{\bf q}_2)Dq_1^2\theta(t_1-t_2)e^{-\bar{D}q_1^2(t_1-t_2)},
\\
G^{(0)}_{B\rho}(12)&=&-\rho_0(2\pi)^d\delta({\bf q}_1+{\bf q}_2)Dq_1^2\theta(t_2-t_1)e^{-\bar{D}q_1^2(t_2-t_1)},
\\
G^{(0)}_{BB}(12)&=&0.
\ee

Notice that $G_{\rho B}(12)$ is retarded in time, $G_{B\rho}(12)$ is advanced, and $G_{BB}(12)=0$. (In fact, all cumulants over only $B$ fields identically vanish: $G_{B\ldots B}(1\ldots n) = 0$.) These results agree with recent work\cite{VCCA} that shows the statistical dynamics of the density of noninteracting Brownian particles can be described by a cubic field theory where the density is the fundamental field.

In the case of Newtonian dynamics, cumulants similarly are given by Eq.~(\ref{eq:cumulant}), but now with
\be
N_{n}=-\frac{p_0^2}{2m^2}\bigg[\sum_{i=1}^{n}q_{i}|t_{i}-t_{0}|\bigg]^2
\ee
and
\be
b(j)= \sum_{i=1}^{n} \frac{{\bf q}_i\cdot {\bf q}_j}{m}\theta (t_{i}-t_{j})(t_i-t_j),
\ee
and where $p_0^2 = m\beta^{-1} = m^2v_0^2$ is the thermal momentum.  Explicitly, the two-point cumulants are
\be
G^{(0)}_{\rho\rho}(12)&=&\rho_0(2\pi)^d\delta({\bf q}_1+{\bf q}_2)e^{-p_0^2q_1^2(t_1-t_2)^2/2m^2},
\\
G^{(0)}_{\rho B}(12)&=&-\rho_0(2\pi)^d\delta({\bf q}_1+{\bf q}_2)\frac{q_1^2}{m}(t_1-t_2)\theta(t_1-t_2)
e^{-p_0^2q_1^2(t_1-t_2)^2/2m^2},
\\
G^{(0)}_{B\rho}(12)&=&-\rho_0(2\pi)^d\delta({\bf q}_1+{\bf q}_2)\frac{q_1^2}{m}(t_2-t_1)\theta(t_2-t_1)
e^{-p_0^2q_1^2(t_1-t_2)^2/2m^2},
\\
G^{(0)}_{BB}(12)&=&0.
\ee

There are some similarities with the Smoluchowski form -- $G_{\rho B}(12)$ is again retarded, $G_{B\rho}(12)$ is advanced, and $G_{BB}(12)$ vanishes -- but the time dependence here is a Gaussian, rather exponential decay.

Importantly, we find that the two-point cumulants (regardless of dynamics) obey the fluctuation-dissipation relation
\be
G_{\rho B}(12) = \theta(t_1-t_2)\beta\frac{\partial}{\partial t_1}G_{\rho\rho}(12)
\label{eq:FDR}
\ee
which holds both order-by-order and for the full cumulant. Additional fluctuation-dissipation relations for $n$-point fields also exist\cite{NDII} and are essential for establishing properties of the cumulants and associated vertex functions.

We next define the vertex function (or matrix inverse) via Dyson's equation
\be
\Gamma_{\alpha\mu}(13)G_{\mu\beta}(32) = \delta_{\alpha\beta}.
\ee
At zeroth order, we have the two point vertices
\be
\Gamma^{(0)}_{B\rho}(12)&=&-\frac{1}{\rho_{0}q_{1}^{2}} \left(\frac{\partial}{\partial t_{1}}+\bar{D}q_{1}^{2}\right)
\delta (t_{1}-t_{2}),
\\
\Gamma^{(0)}_{\rho B}(12)&=&-\frac{1}{\rho_{0}q_{1}^{2}} \left(-\frac{\partial}{\partial t_{1}}+\bar{D}q_{1}^{2}\right)
\delta (t_{1}-t_{2}),
\\
\Gamma^{(0)}_{B B}(12)&=&-\frac{2D}{\rho_{0}q_{1}^{2}}\delta (t_{1}-t_{2}),
\\
\Gamma^{(0)}_{\rho\rho}(12)&=&0
\ee
for Smoluchowski dynamics. The inverses in Newtonian dynamics are considerably more complicated and will be omitted here.

\subsection{First-order solution}
Next, we have the two-point correlation function at first order. The kinetic equation is
\be
G^{(1)}_{\alpha\beta}(12)=G_{\alpha\beta}^{(0)}(12)
+G_{\alpha\gamma}^{(0)}(13)\sigma_{\gamma\delta}(34)G^{(1)}_{\delta\beta}(42),
\label{eq:81}
\ee
where again we sum over repeated subscripts and integrate over repeated argument numbers.

Working out the full solution, we find
\be
G^{(1)}_{\rho\rho}(q,t)= S(q)e^{-\bar{D}\bar{\rho}q^{2}t/S(q)}
\label{eq:70}
\ee
where the static structure factor is given by
\be
S(q)=\frac{\bar{\rho}}{1+\bar{\rho}\beta V(q)}
\label{eq:71}
\ee
and where the bare density $\rho_0$ is everywhere replaced by the first-order revised average density
\be
\bar{\rho} = \frac{\rho_0}{1+\rho_0\beta V(q=0)}.
\label{eq:72}
\ee
We see that again that decay is exponential, but now the relaxation time is inversely proportional to the static structure factor leading to a slowing down near the first structure factor peaks. This is the well-known de Gennes narrowing form\cite{deGennes}.

At this level of  approximation, let us review how we can use these results in practice.

Our theory requires one input---a static structure factor $S(q)$---which is used by Eq.~(\ref{eq:71}) to find the effective potential---also called the pseudopotential---$V(q)$. This in turn updates the density, Eq.~(\ref{eq:72}), and yields the density-density correlation function, Eq.~(\ref{eq:70}). Advanced or retarded response functions $G_{B\rho}(q,t)$ or $G_{\rho B}(q,t)$ could be found, if needed, through the fluctuation-dissipation relation, Eq.~(\ref{eq:FDR}).

At first order, Eq.(\ref{eq:71}) is in the form of the static Ornstein-Zernike relation\cite{hansen}, and we can identify the effective interaction with $c_{D}(q)$, the physical direct correlation function, which is assumed to be known by other means\cite{percus,ashcroft}:
\be
V(q)=-\beta^{-1}c_{D}(q).
\ee

We see, then, that any theoretical or experimentally measured structure factor $S(q)$ or direct correlation function $c_D(q)$ is sufficient to work out the full dynamic results. Already at first order in the perturbation, we have a theory valid for low density fluids and a clear mechanism for going further.

\subsection{The kinetic equation and the memory function: second-order solution}
As we continue expanding in potential, we find that Dyson's equation (in conjunction with the fluctuation-dissipation relation) again gives rise to a kinetic equation governing the full dynamics of the intermediate structure factor, $G_{\rho\rho}(q,t)$. At second-order, we have
\be
\bigg[D_q\frac{\partial^2}{\partial t^2}-A_q\frac{\partial}{\partial t} -\beta^{-1}S^{-1}(q)\bigg]G_{\rho\rho}(q, t) =\int_0^t ds \beta K(q,t-s)\frac{\partial}{\partial s}G_{\rho\rho}(q,s)
\ee
where for Smoluchowski dynamics\cite{SDENE,KEGSD} we have
\be
D_q^{(SD)} = 0, ~~ \text{and} ~~ A_q^{(SD)} = \frac{1}{\beta\bar{\rho}\bar{D}q^2}
\ee
and for Newtonian dynamics\cite{NDII,ULTD} we have
\be
D_q^{(ND)} = \frac{-m\beta}{\beta\bar{\rho}q^2} ~~ \text{and} ~~ A_q^{(ND)} = 0,
\ee
where $\bar{\rho}$ is the density revised to second order, and and where $K(q,t) = -\Gamma_{BB}(q,t)$ is the memory function. These equations are structurally identical to the well-known form derived in MCT\cite{goetze92,goetzerev,goetzebook,das}, however we can already see how this this theory now allows one to continue computing the equation of state $\bar{\rho}$ and the memory function $K(q,t)$ to arbitrary order.

At this order, the effective potential $V(q)$ is the solution to the equation for the static structure factor
\be
S(q) = \frac{1}{1+V(q)-M(q)},
\label{eq:S(q)second}
\ee
where
\be
M(q) = \frac{\pi}{12\eta}\int \frac{d^d{\bf k}}{(2\pi)^d}V(k)S(k)V(|{\bf q}-{\bf k}|)S(|{\bf q}-{\bf k}|),
\ee
and the average density $\bar{\rho}$ is found via
\be
\rho_0 = \bar{\rho}\bigg[V(q=0) -\frac{1}{2\bar{\rho}}\int\frac{d^d{\bf k}}{(2\pi)^d} V^2(k)S(k)\bigg].
\ee

Normalizing the intermediate structure factor,
\be
F(q,t) = G_{\rho\rho}(q,t)/G_{\rho\rho}(q,t=0) = G_{\rho\rho}(q,t)/S(q),
\ee
switching from density $\bar{\rho}$ to packing fraction,
\be
\eta = \frac{\pi\bar{\rho}\sigma^3}{6},
\ee
and moving to dimensionless quantities,
\be
q\sigma &\rightarrow& q\\
p_0t/m\sigma \rightarrow t &\textrm{or}&
\bar{D}t/\sigma^2 \rightarrow t\\
S(q)/\bar{\rho} &\rightarrow& S(q)\\
\bar{\rho}\beta V(q) &\rightarrow& V(q)\\
\beta^2\bar{\rho}K(q,t) &\rightarrow& K(q,t),
\ee
we have
\be
\frac{\partial}{\partial t}F(q,t)
&=&-q^2S^{-1}(q)F(q,t)+q^2\int_0^{t} ds K(q,t-s) \frac{\partial}{\partial s}F(q,s)
\label{eq:kinetic}
\ee
for the Smoluchowski case and
\be
\frac{\partial^2}{\partial t^2}F(q,t)
&=&-q^2S^{-1}(q)F(q,t)+q^2\int_0^t ds K(q,t-s) \frac{\partial}{\partial s}F(q,s)
\ee
for the Newtonian case.

The full memory function was derived to second order in the potential in Ref. \onlinecite{KEGSD} for Smoluchowski dynamics and in Ref. \onlinecite{NDII} for Newtonian dynamics (in the large wavelength limit only). As discussed there, correcting the vertex functions to higher order will add derivative terms which influence short-time, but not long-time dynamics; consequently, we will use their zeroth-order approximations in this work to simplify the equations.

Summarizing here, the memory function is
\be
K(q,t) = K^{(s)}(q,t) + K^{(c)}(q,t)
\label{eq:NoV}
\ee
where the self contribution is given by
\be
K^{(s)}(q,t)&=&\frac{\pi}{12\eta}\int \frac{d^d {\bf k}}{(2\pi)^d}
\bigg[V^2(k)S(k)\tilde{F}(k,t)F^{(0)}(|{\bf q-k}|,t)
\nonumber\\
&&+V^2(|{\bf q-k}|)S(|{\bf q-k}|)F^{(0)}(k,t)\tilde{F}(|{\bf q-k}|,t)\bigg]
\ee
and where the collective contribution is given by
\be
K^{(c)}(q,t)&=&\frac{\pi}{12\eta}\int \frac{d^d {\bf k}}{(2\pi)^d}
V(k)S(k)V(|{\bf q-k}|)S(|{\bf q-k}|)\bar{F}(k,t)\bar{F}(|{\bf q-k}|,t).
\ee
Note that these terms do not depend on the full propagator $F(q,t)$ but on the dressed propagators $\bar{F}(q,t)$ and $\tilde{F}(q,t)$,
\be
\bar{F}(q,t) &=& \bar{G}_{\rho\rho}(q,t)/\bar{G}_{\rho\rho}(q,t=0)\\
\tilde{F}(q,t) &=& \tilde{G}_{\rho\rho}(q,t)/\tilde{G}_{\rho\rho}(q,t=0)
\ee
and the noninteracting propagator
\be
F^{(0)}(q,t) &=& G^{(0)}_{\rho\rho}(q,t)/G^{(0)}_{\rho\rho}(q,t=0)
\ee
which will decay to zero at long times regardless of the dynamics.

The dressed propagators arise naturally in the derivation and behave much like the normal correlation function, $F(q,t)$. Most importantly, the dressed propagators $\bar{G}_{\alpha\beta}$ and $\tilde{G}_{\alpha\beta}$ are themselves subject to fluctuation-dissipation relations.

In the frequency domain, the cumulants obey the matrix relations
\be
\bar{G}_{\alpha\beta}(12)&=&\frac{1}{2}\bigg[G^{(0)}_{\alpha\gamma}(13)\sigma_{\gamma\delta}(34)G_{\delta\beta}(42) + G_{\alpha\gamma}(13)\sigma_{\gamma\delta}(34)G^{(0)}_{\delta\beta}(42)\bigg]
\ee
and
\be
\tilde{G}_{\alpha\beta}(12)&=& G^{(0)}_{\alpha\gamma}(13)\sigma_{\gamma\delta}(34)
G_{\delta\epsilon}(45)\sigma_{\epsilon\zeta}(56)G^{(0)}_{\zeta\beta}(62),
\ee
and in the time domain, the normalized $\rho\rho$-components become
\be
\bar{F}(q,t) &=& F^{(0)}(q,t) + q^2\int_0^t ds~ F^{(0)}(q,t-s)F(q,s)
\ee
and
\be
\tilde{F}(q,t) &=& F^{(0)}(q,t)\bigg[1+q^2|t|\bigg] + q^4\int_0^t ds~ |t-s|F^{(0)}(q,t-s)F(q,s).
\ee

Importantly, both functions decay quicker than $F(q,t)$ at short times, but approach the full correlation function at long times:
\be
\bar{F}(q,t) &\rightarrow& F(q,t), ~~t \gg 1\\
\tilde{F}(q,t) &\rightarrow& F(q,t), ~~t \gg 1.
\ee

In fact, we can use the long-time limits to form an approximate memory function
\be
K^{(LT)}(q,t) = \frac{\pi}{12\eta}\int \frac{d^d {\bf k}}{(2\pi)^d}
V(k) S(k) V(|{\bf q}-{\bf k}|) S(|{\bf q}-{\bf k}|) F(k,t) F(|{\bf q}-{\bf k}|,t)
\label{eq:LT}
\ee
where we have replaced $\bar{F}(q,t)$ and $\tilde{F}(q,t)$ with $F(q,t)$ and where we have dropped the self contribution which will be subdominant to the collective contribution at long time due to its dependence on the noninteracting cumulant $F^{(0)}(q,t)$.

This longtime approximate form is strikingly similar to the mode-coupling theory memory function, Eq.~(\ref{eq:MCT}); it is quadratic in the potential and quadratic in the density-density time correlation function, though importantly Eq.~(\ref{eq:LT}) has been derived from first principles by a controlled, systematic approximation.

\subsection{Static Solution at second-order}
As shown in Ref. \onlinecite{spyridis}, we can write down an equation for the the nonergodicity factor, $f_q$, in terms of static quantities only:
\be
\frac{f_q}{1-f_q} = \mathcal{F}[f_q]
\ee
where
\be
\mathcal{F}[f_q] = \frac{\pi}{12\eta}\int \frac{d^d {\bf k}}{(2\pi)^d}
V(k) S(k) V(|{\bf q}-{\bf k}|) S(|{\bf q}-{\bf k}|) f_k f_{|{\bf q}-{\bf k}|}.
\ee
We may rearrange to find
\be
f_q = \frac{\mathcal{F}[f_Q]}{1+\mathcal{F}[f_Q]}
\label{eq:static}
\ee
which can be solved by iteration. (An initial guess for $f_q$ yields an initial value for $\mathcal{F}[f_q]$. This, in turn revises the estimate for $f_q$ by Eq.~(\ref{eq:static}) and the cycle repeats.)

In the liquid phase, only the trivial solution $f_q = 0$ for all $q$ is supported, but at the transition density, the solution bifurcates, allowing the possibility of a discontinuous jumps to a second, positive result.

\subsection{Asymptotic expansion prediction}
One of the most important predictions of mode-coupling theory is that the relaxation of the system through the $\beta$-regime is governed by a set of master equations and that these master equations predict power-law decay into and out of the plateau. In addition, the exponents are predicted to be universal properties of the system and the time scales associated with the $\alpha$- and $\beta$-regimes themselves are predicted to diverge as power-laws.

Recent work shows that the formulation of particle dynamics introduced here at second order in the potential can be treated identically as in MCT\cite{spyridis}; our system is expected to display the same power-law decay structure, though the values of the associated parameters may be different. Let us review the predictions.

First, though the behavior near the transition is universal, the location of the transition is system-specific. For this reason, we change from temperature $T$ (or packing fraction $\eta$) to the system-independent separation parameter $\epsilon = (T^*-T)/T^*$ (or $\epsilon = (\eta-\eta*)/\eta^*$). Note that $\epsilon$ is negative when we are in the ergodic regime ($T > T^*$ or $\eta < \eta*$) and it is positive when we are in the nonergodic regime ($T < T^*$ or $\eta > \eta*$).

In the $\beta$-regime, we can generically describe the behavior as a small deviation from the plateau,
\be
F(q,t) = f_q^* + h_qG(t/\tau_{\beta}),
\ee
where $f_q^*$ is the nonergodicity factor at the critical density, $h_q$ is the so-called critical amplitude, and $G(t/\tau_{\beta})$ is the master function,
\begin{equation}
   G(t/\tau_{\beta}) = \left\{
     \begin{array}{lr}
       \sqrt{|\epsilon|}g_{\pm}(t/\tau_{\beta}) & : \epsilon \neq 0 \\
       (t/\tau_0)^{-a} & : \epsilon = 0
     \end{array}
   \right.
\end{equation}
where the plus or minus subscript designates $\epsilon > 0$ or $\epsilon < 0$ respectively. Here, $0 < a < 1/2$ is the first critical exponent and $g_{\pm}(t/\tau_{\beta})$ has limiting forms
\be
g_{\pm}(t/\tau_{\beta}) = (t/\tau_{\beta})^{-a} &:& \tau_0 \ll t \ll \tau_{\beta} ; \epsilon \neq 0\\
g_{+}(t/\tau_{\beta}) = 1/\sqrt{1-\lambda}      &:& \tau_\beta \ll t \ll \tau_{\alpha} ; \epsilon > 0 \\
g_{-}(t/\tau_{\beta}) = -(t/\tau_{\beta})^{b}   &:& \tau_\beta \ll t \ll \tau_{\alpha} ; \epsilon < 0
\ee
where $0 < b < 1$ is the second critical parameter and $\lambda$ is the exponent parameter relating $a$ and $b$,
\be
\lambda = \frac{\Gamma(1-a)^2}{\Gamma(1-2a)} = \frac{\Gamma(1+b)^2}{\Gamma(1+2b)},
\label{eq:lambda}
\ee
which is constrained to be $1/2 \leq \lambda \leq 1$.

The two time scales diverge as
\be
\tau_{\beta} = \frac{\tau_0}{|\epsilon|^{1/2a}}
\ee
and
\be
\tau_{\alpha} = \frac{\tau_0}{|\epsilon|^{\gamma}} = \frac{\tau_{\beta}}{|\epsilon|^{1/2b}}
\label{eq:taualpha}
\ee
where $\gamma = 1/2a+1/2b$. Therefore, we see that the ratio of the time scales diverges on approaching the transition from the ergodic side as
\be
\frac{\tau_{\alpha}}{\tau_{\beta}} = \frac{1}{|\epsilon|^{1/2b}}.
\ee

Using these time scales, we can rewrite our forms for the $\beta$-regime in the nonergodic phase as
\begin{equation}
   F(q,t) -f_q^* = \left\{
     \begin{array}{ll}
        h_q(t/\tau_0)^{-a}           & : \tau_0 \ll t \ll \tau_{\beta}; \epsilon \geq 0 \\
        h_q\sqrt{\epsilon/(1-\lambda)} & : t \rightarrow \infty; \epsilon \geq 0
     \end{array}
   \right.
\label{eq:nonergodicity}
\end{equation}
and in the ergodic phase as
\begin{equation}
   F(q,t) -f_q^* = \left\{
     \begin{array}{ll}
        h_q(t/\tau_0)^{-a}       & : \tau_0 \ll t \ll \tau_{\beta}; \epsilon < 0 \\
        -h_q(t/\tau_{\alpha})^b    & : \tau_{\beta} \ll t \ll \tau_{\alpha}; \epsilon < 0
     \end{array}
   \right..
\end{equation}
Thus, we see that in the approach to the plateau, both the ergodic and nonergodic systems behave identically; it is only at long-times when the system decays out of the plateau (or not) that a distinction can be made.

In the nonergodic phase, no simple model for the final relaxation to zero through the $\alpha$-regime falls out of the asymptotic expansion. The initial decay out of the plateau -- where the $b$-exponent power law form holds -- is termed the von Schweidler relaxation and one might assume this form would hold all the way to zero. It is believed, however, that there is a complex interplay of relaxation on different length scales such that the power-laws at different wavenumbers become superimposed\footnote{One explanation for the stretched-exponential nature of the $\alpha$-regime is that there are many domains within the fluid each relaxing at their own rate; the observed behavior at large-scale is an effective relaxation that may in fact be quite different from the small-scale behavior within a given domain. This idea, called dynamical heterogeneity, has been seen in both experiment and simulation\cite{goetze92, adamgibbs, sillescu,glotzer,ediger00}, though is not predicted by mode-coupling theory. Rather than look at the 2-point density-density time correlation function we study in this work, it has been proposed that the quantity to study is the 4-point correlation function relating shifted products of the density, $\rho({\bf r}, t)\rho({\bf r}+ {\bf r_0}, t + \tau)$: ${C_{4}({\bf r},t) =\langle \delta (\rho ({\bf 0},0)\rho ({\bf 0}+{\bf r}_{0},\tau ))
\delta (\rho ({\bf r},t)\rho ({\bf r}+{\bf r}_{0},t+\tau ))\rangle}$. This correlation function should scale with a length $\ell$ which diverges at the ergodic-nonergodic transition\cite{berthier04, berthier05, BB, BBMR, garrahan}. While we do not address it in this paper, our theory has all the machinery necessary to study 4-point quantities such as this one and could shed some light on the nature of the diverging length scale.}. Experimentally, this so-called time-temperature superposition is remarkably well modeled as a stretched-exponential of the Kohlrausch-Wiliams-Watts form\cite{kohlrausch, ww}
\be
F(q,t) = A_q\exp[-(t/\tau_q)^{\beta_q}],
\label{eq:stretchedexp}
\ee
where $\beta_q \leq 1$ and where $A_q \leq f_q$ allows for the fact that the exponential behavior need not take over until some time after the initial decay away from the plateau. The effective time constant $\tau_q$ defined here is wavenumber-dependent (due to the mixing described above), but will be proportional to and of the same order as $\tau_{\alpha}$.

It has been shown under MCT\cite{Fuchs94} that at large $q$ the stretching exponent can be related to the von Schweidler decay such that
\be
\lim_{q\rightarrow\infty}\beta_q = b.
\ee
While we will fit to the stretched exponential form here, we will not explore correspondence in this limit, a historically difficult endeavor.

As $\tau_q$ is proportional to $\tau_{\alpha}$, scaling time in the intermediate structure factor by $\tau_{\alpha}$ causes the longtime behavior of $F(q,t)$ at all densities near the transition to collapse onto one curve from the von Scheweidler decay out of the $\beta$-regime through the whole of the $\alpha$-regime.

\section{Solution for hard spheres obeying Smoluchowski dynamics}
\label{sec:solution}
\subsection{Defining the system}
To this point, we have kept the discussion general as this theory is applicable to a wide variety of systems. However, let us now look at a concrete example and solve the kinetic equation for a system of hard spheres obeying Smoluchowski dynamics. We choose this system first as it is a simple, but non-trivial system that shows all of the dynamic features predicted thus far. In addition, working with hard spheres allows us a straightforward comparison with simulation and experiment; as will be discussed in later sections, an appropriate analogue to hard spheres is a colloid suspension\cite{pusey86} and there is detailed work on such systems against which we may compare our results\cite{pusey87,phan,vanmegen98,cheng,goetze99,brambilla}.

\begin{figure}[p]
\begin{subfigure}[b]{0.75\columnwidth}
    \includegraphics[width=\textwidth]{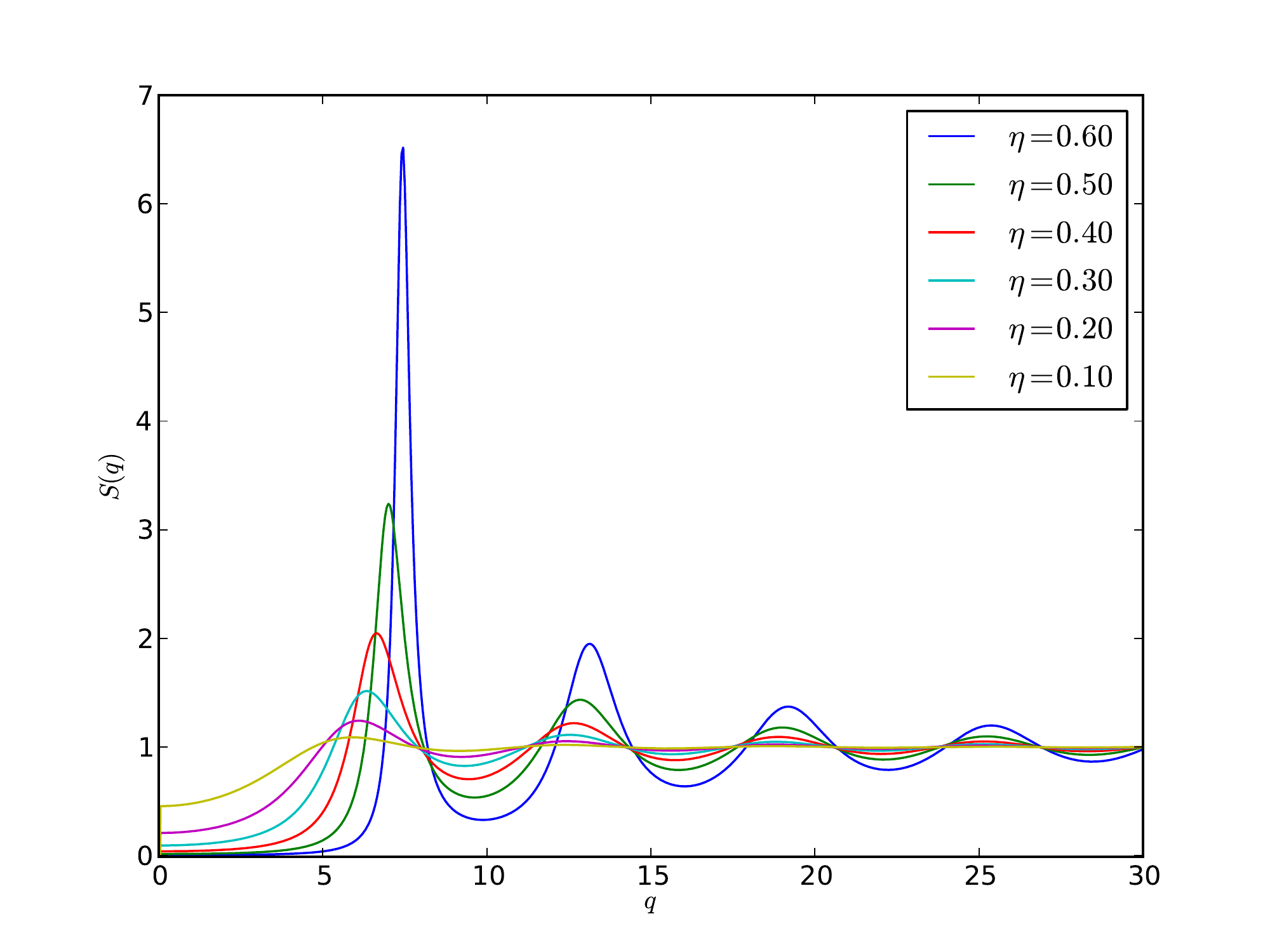}
    \caption{}
\end{subfigure}
\begin{subfigure}[b]{0.75\columnwidth}
    \includegraphics[width=\textwidth]{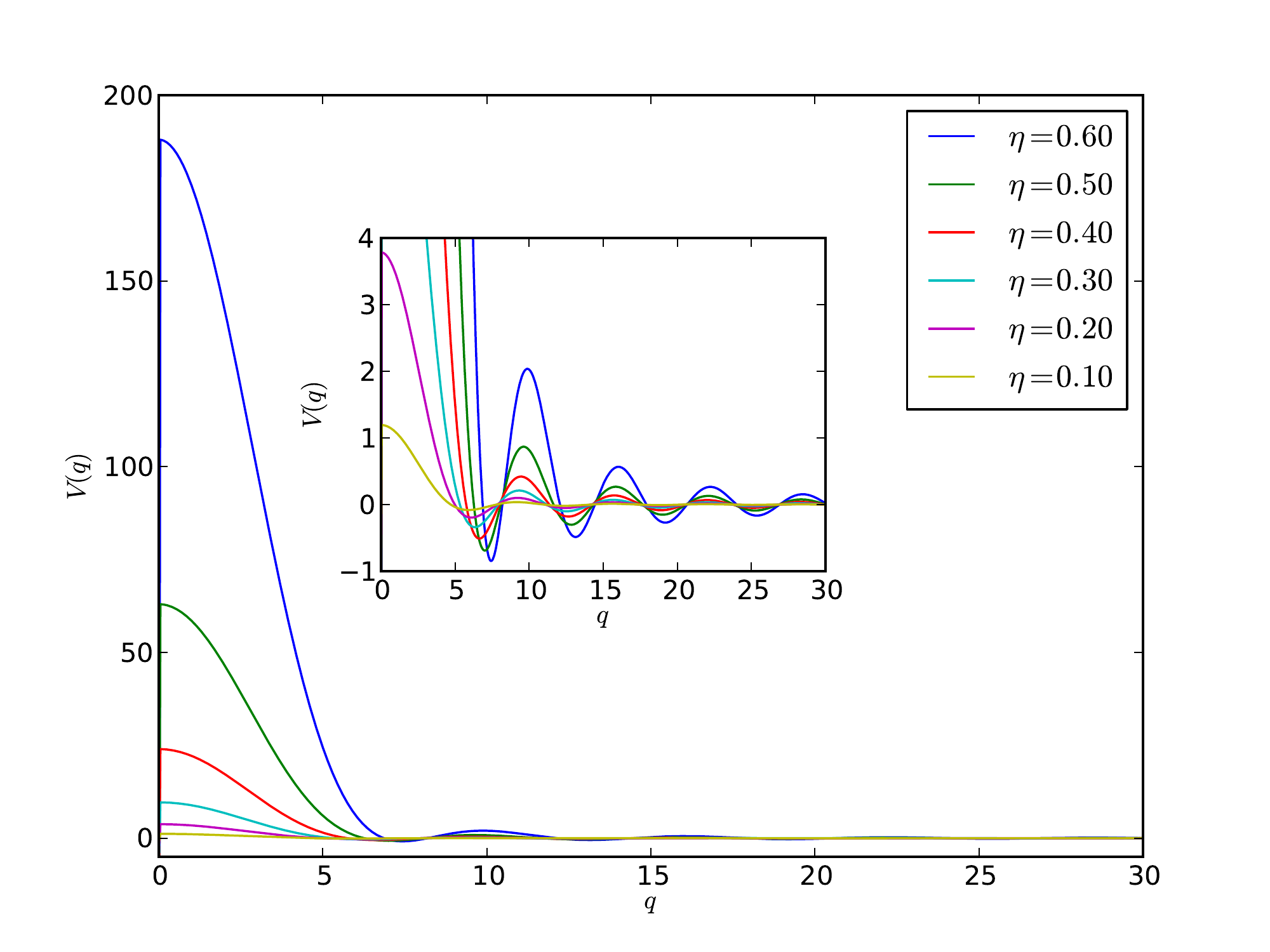}
    \caption{}
\end{subfigure}
\caption{(a) The dimensionless structure factor $S(q)$ and (b) the Percus-Yevick effective potential $V(q)=-c_D(q)$ for hard spheres are plotted for several packing fractions, $\eta$. The inset in (b) shows the detail of $V(q)$ about zero.}
\label{fig:S(q)}
\end{figure}

Recall that the theory requires one input -- a static structure factor. As is common with hard spheres, we will use the Ornstein-Zernike equation for the static structure factor\cite{hansen}, written here in dimensionless variables as
\be
S(q) = \frac{1}{1-c_D(q)},
\ee
with the Percus-Yevick approximation for the direct correlation function, $c_D(q)$\cite{percus,ashcroft}. At first order in our theory, the effective potential (pseudopotential) is given by
\be
V(q) = S^{-1}(q) - 1
\ee
which is equivalent to the substitution $V(q) = -c_D(q)$. Plots of $S(q)$ and $V(q)$ are shown in Fig. \ref{fig:S(q)}.

Note that in the case of hard spheres, temperature does not appear at all; the only control parameter is the packing fraction density, $\eta$.

As discussed above, we may update our effective potential to second order through Eq.~({\ref{eq:S(q)second}). The effect of including higher-order terms in the pseudopotential, however, causes only a moderate change in the position of the transition as determined by the static solution. While a full investigation of this effect should eventually be done, we will instead here stick with the first-order solution; at this order, both $S(q)$ and $V(q)$ match the commonly-used Percus-Yevick forms, making comparison to other work simpler.

As an aside, we will however note that under the approximation of ignoring vertex corrections, it is possible to sum the series to all orders. Even in this extreme limit, there is little change to the transition density\cite{ULTD}. Full discussion of these results will be addressed elsewhere.

\subsection{Numerical solution}
The kinetic equation, Eq.~(\ref{eq:kinetic}), does not have an analytic solution for general $S(q)$, so we must instead solve it numerically to find the full function $F(q,t) = G_{\rho\rho}(q,t)/S(q)$ subject to the initial conditions $F(q,t=0) = 1$ and $\partial F(q,t)/\partial t|_{t=0} = 0$. The full details of our numerical algorithm are given in Appendix \ref{app:numerical}.

\begin{figure}[p]
\begin{subfigure}{\columnwidth}
    \includegraphics[width=\textwidth]{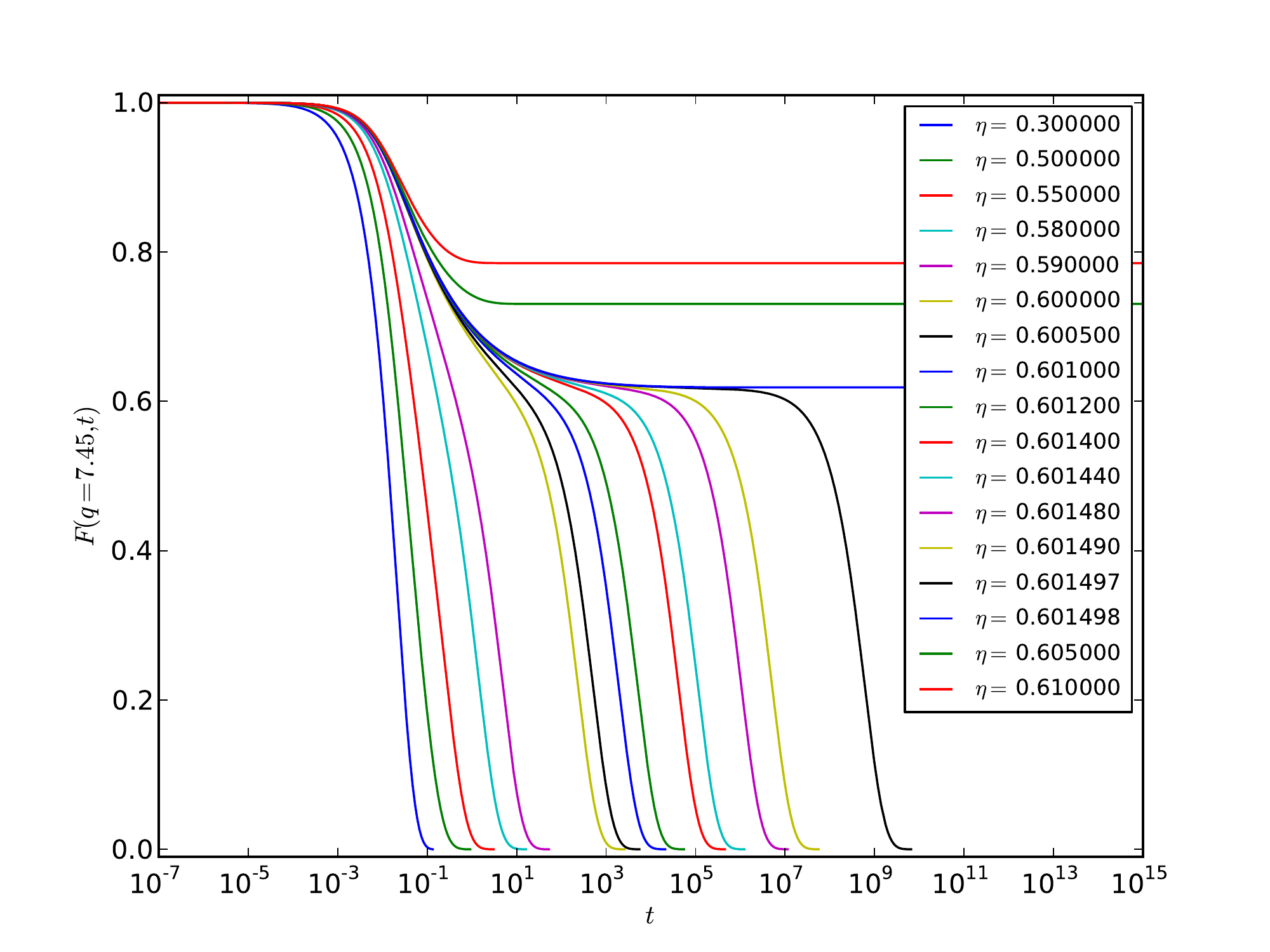}
\end{subfigure}
\caption{$F(q,t)$ is shown as a function of time at fixed wavenumber $q=7.45$ near the first structure factor maximum for select values of packing fraction $\eta$. As the system approaches the transition density of $\eta^*=0.60149761$ from below, there is a dramatic slowing down with the system remaining trapped near the plateau in the $\beta$-regime over many decades. Above the transition, the system becomes ergodic and remains locked in the plateau at long times.}
\label{fig:F(q,t)}
\end{figure}

First, we consider the solution using the longtime limit form of the memory function, Eq.~(\ref{eq:LT}). (Select densities are plotted in Fig. \ref{fig:F(q,t)} at fixed wavenumber.) The intermediate structure factor shows a slower relaxation compared to both the zeroth- and first-order solutions at all densities, and the full two-step decay emergences as the packing fraction rises above $\eta\approx 0.60$. The plateau which appears rapidly grows longer with increasing density until we find the transition to a nonergodic phase for $\eta > 0.601497$. The densities probed here represent decay over more than a dozen decades of time.

Next, we may repeat our solution, but with the more complete, no-vertex correction form of the second-order memory function given in Eq.~(\ref{eq:NoV}). In this case, we must keep track not only of $F(q,t)$, but also the dressed propagators, $\bar{F}(q,t)$ and $\tilde{F}(q,t)$. As discussed in the theory section, both these functions decay faster than $F(q,t)$ at short times, but tend toward the undressed $F(q,t)$ at long times. (Plots of the dressed propagators are shown in Fig. \ref{fig:dressed} in Appendix \ref{app:dressed}.) The solution in this case is identical except for an overall shift in the short time scale, $\tau_0\rightarrow\tau_0^{\prime}$; the two-step decay has the same form and the the ergodic-nonergodic transition occurs at the same packing fraction. We will therefore restrict numerical results discussed for the rest of the paper only to the solution using the longtime form of the memory function.

Numerically solving the kinetic equation for this system represents an important step in the study of the glass-transition problem: this theory is the first outside of mode-coupling theory to show a full dynamic solution with both the two-step decay and an ergodic-nonergodic transition. As will be shown in the following section, this solution reproduces features seen in experiment and simulation and exhibits functional forms and scalings of the same type as MCT. It bares repeating, however, that this form of the theory and the assumptions which go into the numerical solution are better motivated than those of MCT. Additionally, the clear mechanism for instituting corrections will allow future solutions to be carried out at higher order in the expansion of the effective potential, exploring how the physics evolve order-by-order.

\subsection{Fitting}
In order to compare our results to the asymptotic predictions, we need to perform \textit{coordinated} fits looking both at how the predicted forms model the intermediate structure factor and how the extracted fit parameters scale with density. The chief difficulty in performing these fits comes from determining the regions of the data where each predicted form is applicable; several of the parameters change quickly with modest adjustments to the fit domain. This is a well-known challenge within MCT\cite{goetze99,weysser} and for fitting power law forms in general where one typically requires many decades of data to be confident in the results.

Our approach, therefore, will not be to attempt a direct fit to $F(q,t)$ on our first pass, but instead to deduce the values of the critical amplitude and critical exponents indirectly; a fit of $\tau_{\alpha}$ as a function of density will yield critical exponent $\gamma$ which can be used to find $a$ and $b$. On second pass, we will then return to the na\"{i}ve approach of direct fitting. By adhering to a set of criteria qualifying the goodness-of-fit, we show that power-law exponents and associated parameters can be extracted directly and that they match those obtained in the above methods despite having only a few decades of data to work with in each domain.

By finding fit values in these two independent ways, we build confidence in our result.

\subsubsection{The nonergodicity factor, $f_q$}
First, let us look at the nonergodicity factor $f_q$ and investigate how it scales with density.

Recall that $f_q$ had previously been found using the static equation, Eq.~(\ref{eq:static})\cite{spyridis}. We expect that extracting the nonergodicity factor directly from our results the longtime limit,
\be
\lim_{t\rightarrow\infty} F(q,t) = f_q,
\ee
should yield the same results and in fact find negligible difference, verifying the reliability of the static solution. Several nonergodicity factors are shown in Fig \ref{fig:fq}.

\begin{figure}[p]
\includegraphics[width=\columnwidth]{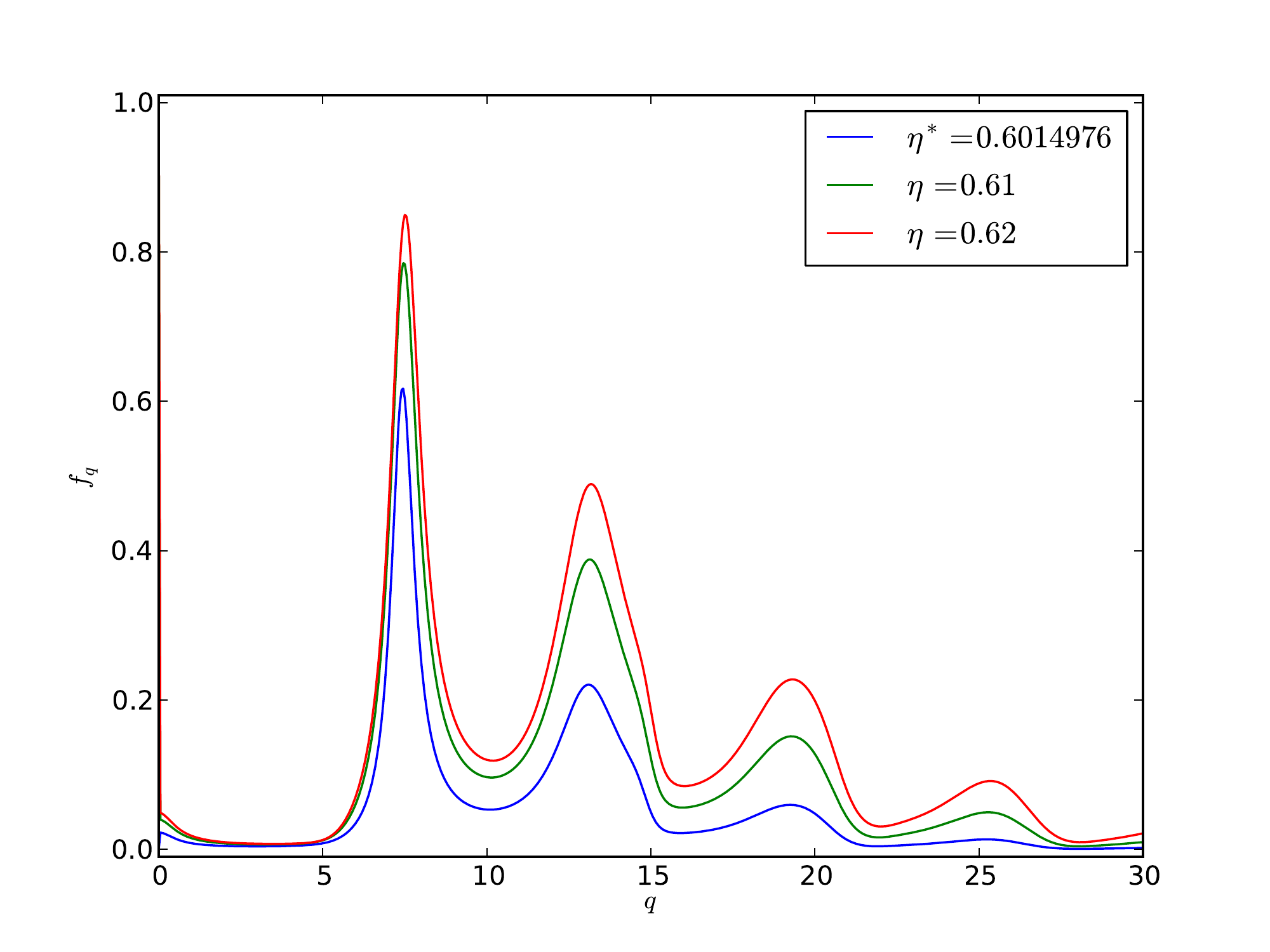}
\caption{A plot of the nonergodicity factor $f_q$ at the transition density, $\eta^*=0.60149761$ along with two curves at slightly higher density. Note that the positions of the peaks correspond well with the positions of the peaks of the static structure factor at the same packing fraction density and that the amplitude grows moderately, though not uniformly, as packing fraction increases.}
\label{fig:fq}
\end{figure}

From Eq.~(\ref{eq:nonergodicity}), we have that at fixed $q$ the nonergodicity factor scales with density as
\be
f_q = f_q^* + h_q\sqrt{\frac{\epsilon}{1-\lambda}}=f_q^* + h_q\sqrt{\frac{\eta-\eta^*}{\eta^*(1-\lambda)}}.
\label{eq:fqfit}
\ee
We fit this form over a wide range of wavenumbers and plot select results in Fig. \ref{fig:fqfits}. Each fit yields a comparable value for the transition density of $\eta^* = 0.60149761(10)$ where the uncertainty comes from the small spread in values over the different fits\footnote{A note on uncertainties: Within this paper, we will quote uncertainties estimated from the spread of values found when performing the same fit over related sets of data. This spread is generally small,---signaling that the fitting forms are indeed universal, at least over the restricted sets of the fits,---but it ignores the systematic uncertainty associated with our specific choices for numerical discretization. We find that there is a large phase space of choices for integration method, step size, array length, etc. that yield \textit{good} solutions that are qualitatively identical, but do show some very small quantitative differences between each other. In the absence of any \textit{a priori} criteria to judge one set of numerical parameters better than another, we have made the most pragmatic choices. The result is that the spread in the quoted statistical uncertainty is small once we define our {``numerical system"} but that these uncertainties are perhaps a bit smaller than the spread in values over \textit{all} such numerical systems for this theory. To give one example, the transition density $\eta^* = 0.60161(2)$ of Ref. \onlinecite{spyridis} (Fig. 9 therein) was found via an iteration technique applied to the static equation and differs by more than one standard deviation from the transition density  $\eta^* = 0.60149761(10)$ found here from the full forward-stepping dynamic solution.}. Coupling these fits with our determination of $\lambda$ in the next section, we extract the critical amplitude $h_q$ which is plotted in Fig. \ref{fig:hq}.

\begin{figure}[p]
\includegraphics[width=\columnwidth]{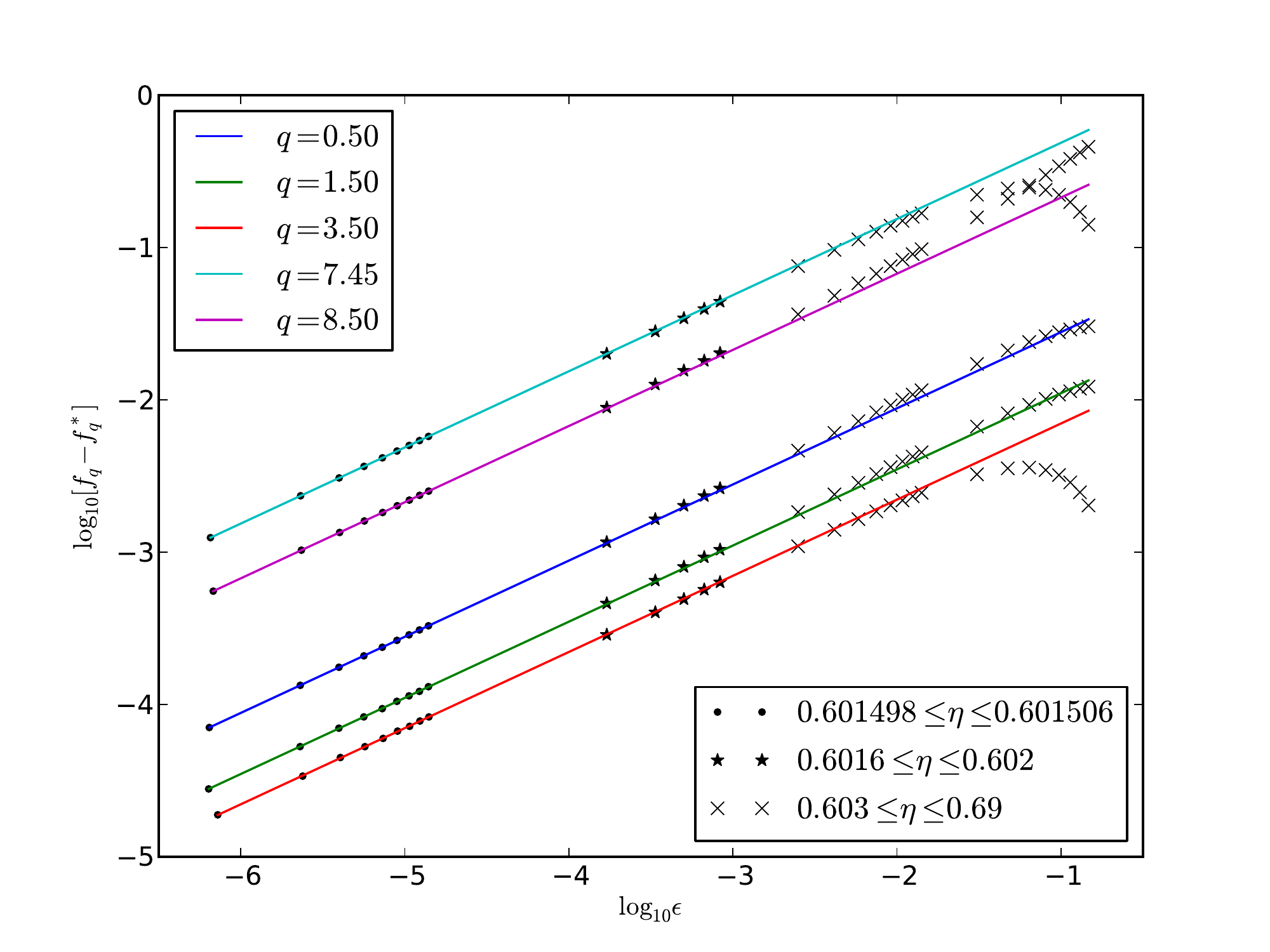}
\caption{Fits of the nonergodicity factor to Eq.~(\ref{eq:fqfit}) for select values of wavenumber are plotted on a loglog scale such that each curve is linear with slope 1/2. Only points for $\eta \leq 0.601506$ ($\epsilon \leq 2\times10^{-5}$) were used in the fit, though the fit function is extended through $\eta = 0.69$ to show where the data depart from the model. For all wavenumbers, agreement is good to $\eta\approx 0.602$ ($\epsilon \approx 10^{-3}$).}
\label{fig:fqfits}
\end{figure}

\begin{figure}[p]
\includegraphics[width=\columnwidth]{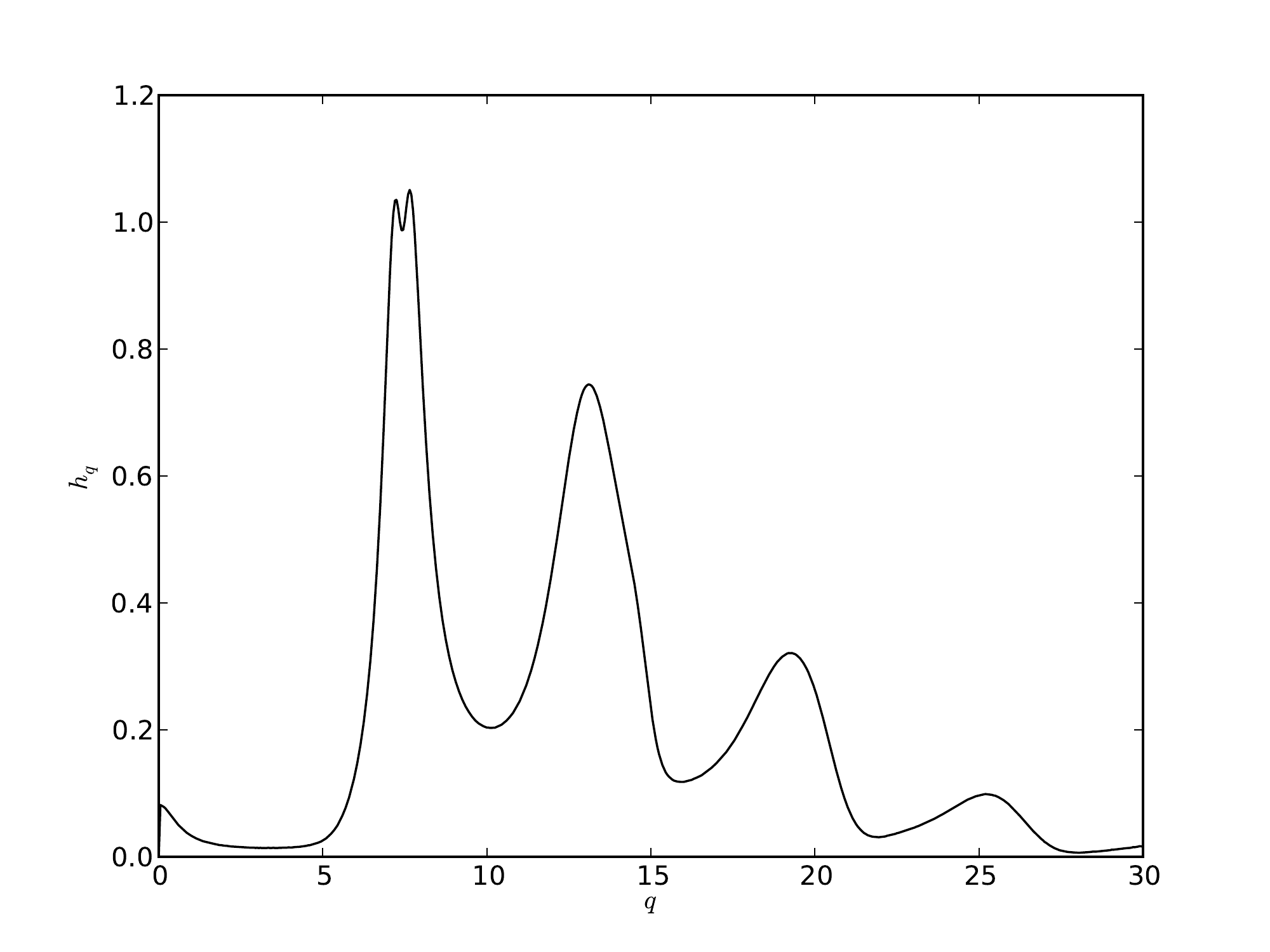}
\caption{The critical amplitude $h_q$ is computed from from fits to the nonergodicity factor, Eq.~(\ref{eq:fqfit}). Note that the peaks of $h_q$ correspond closely to the peaks in $S(q)$ and $f_q$.}
\label{fig:hq}
\end{figure}

\subsubsection{The $\alpha$-relaxation time scale, $\tau_{\alpha}\propto\tau_q$}
Next, we look at the $\alpha$-relaxation in the ergodic phase. As described above, one can scale the time variable of the intermediate structure factor by $\tau_{\alpha}$ and collapse the function onto one master curve at long times.

While it is possible to extract $\tau_{\alpha}$ from the data, we need not find that time specifically. Any time proportional to $\tau_{\alpha}$ will show the same scaling and $\tau_q$---the effective time constant of the stretched exponential fit of the $\alpha$-relaxation, Eq.~(\ref{eq:stretchedexp})---is the easiest to extract. Rewriting Eq.~(\ref{eq:taualpha}) here, we have
\be
\tau_q = \frac{\theta_0}{|\epsilon|^{\gamma}}=\frac{\theta_0\eta^*}{(\eta^*-\eta)^{\gamma}}
\label{eq:tauq}
\ee
where $\gamma = 1/2a + 1/2b$ and where $\theta_0$ is a fitting parameter proportional to the microscopic time scale, $\tau_0$.

Fitting several $\alpha$-relaxation decays at different densities, we extract $\tau_q$ and plot the collapsed function $F(q,t/\tau_q)$ in Fig. \ref{fig:longtimescaling}; all curves collapse to one from the von Schweidler decay through the stretched exponential decay to zero.

\begin{figure}[p]
\includegraphics[width=\columnwidth]{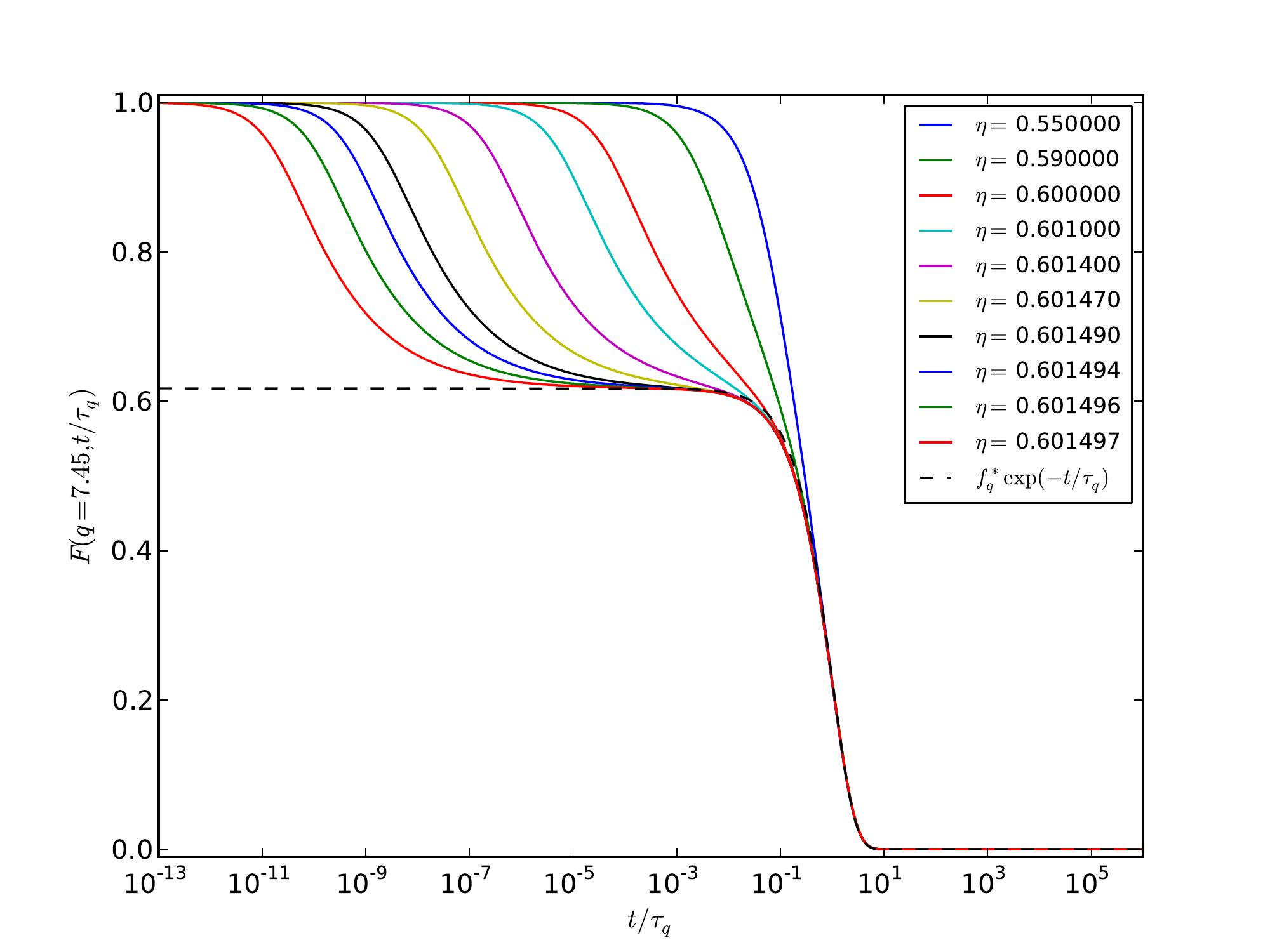}
\caption{The intermediate structure factor is plotted for several densities against scaled time $t/\tau_q$ where $\tau_q$ is found as the effective time constant of the stretched exponential, Eq.~(\ref{eq:stretchedexp}). As $\tau_q$ is proportional to $\tau_{\alpha}$, the longtime behavior---from the von Schweidler decay out of the plateau through the $\alpha$-relaxation to zero---collapses onto one curve under such a transformation. We also plot an exponential function of the form $f_q\exp(-t/\tau_q)$ for comparison showing that the $\alpha$-relaxation is well approximated by a stretched exponential with $\beta$ very close to one.}
\label{fig:longtimescaling}
\end{figure}

Next, we show the explicit scaling of $\tau_q$ as a function of density for select values of wavenumber in Fig. \ref{fig:taualpha}. Power law fits to this data find $\gamma = 1.887(4)$ where the uncertainty is given by the small spread in fit values at different wavenumbers.

\begin{figure}[p]
\includegraphics[width=\columnwidth]{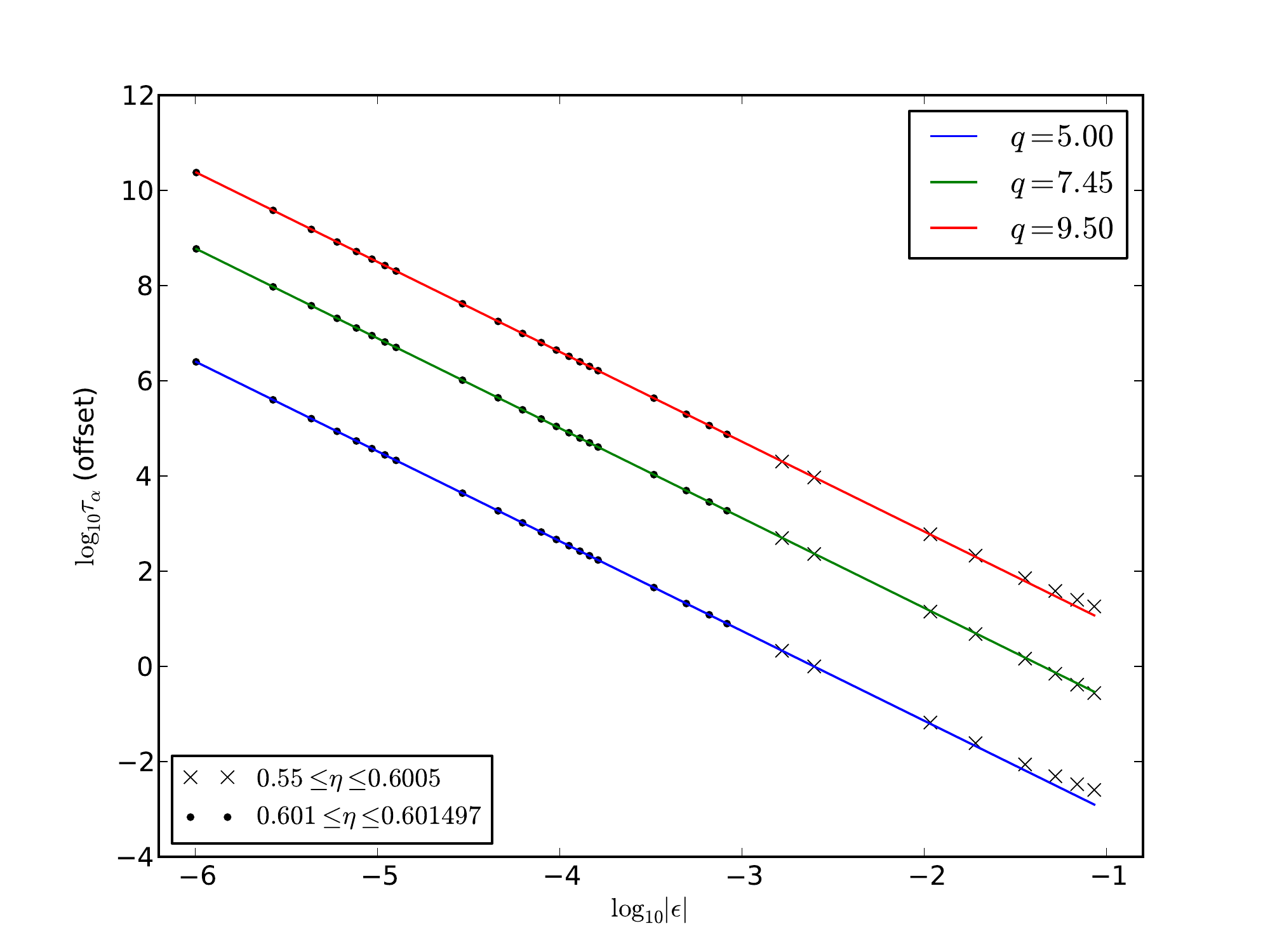}
\caption{Fits of the time scale $\tau_q$ to Eq.~(\ref{eq:tauq}) for select values of wavenumber are shown on a log-log scale such that each curve is linear with slope $\gamma = 1/2a+1/2b = 1.887$. Only points for $\eta \geq 0.601$ ($|\epsilon| \leq 10^{-3}$) were used in the fit, though the fit function is extended down to $\eta = 0.55$ to show where the data depart from the model. (Note that the curve and data for $q=9.50$ has been artificially shifted up by 2 units and $q=5.00$ shifted down by 2 units in order to prevent data sets from overlapping on the plot. Scaling is unaffected.)}
\label{fig:taualpha}
\end{figure}

We now have one equation -- $\gamma = 1/2a + 1/2b$ -- with two unknowns. If we combine this result with the constraint given by $\lambda$, Eq.~(\ref{eq:lambda}), we identify a unique set of exponents and find
\begin{equation}
\begin{array}{ccc}
a=0.375(3),~ & b=0.887(4),~ & \lambda=0.5587(18).
\end{array}
\end{equation}

\subsubsection{Verifying the predicted exponents for the intermediate structure factor, F(q,t)}
In the ergodic phase, we fit to the form
\begin{equation}
   F(q,t) = \left\{
     \begin{array}{ll}
        f_a + h_q(t/\tau_0)^{-a}        & : \tau_0 \ll t \ll \tau_{\beta}\\
        f_b - h_q(t/\tau_{\alpha})^b    & : \tau_{\beta} \ll t \ll \tau_{\alpha}
     \end{array}
   \right.,
   \label{eq:fit1}
\end{equation}
where we fix $a$, $b$, and $h_q$, but allow the other parameters to vary. Note that we have allowed the plateau value -- $f_a$ in the first line and $f_b$ in the second -- to be fit independently at the early and late times even though we expect the value to be the same in both fits. As we will see below, this constraint is recovered naturally when the appropriate domain for the fit is selected.

In the nonergodic phase, we fit to
\be
F(q,t) = f_a + h_q(t/\tau_0)^{-a} ~~: \tau_0 \ll t \ll \tau_{\beta}
\label{eq:fit2}
\ee
where again we keep $h_q$ and $a$ fixed and expect $f_a$ to approach the critical nonergodicity factor as the fit domain choice improves.

As discussed above, picking domain cuts by eye can lead to great variation, but we institute a set of criteria to determine the optimal domain.
\begin{itemize}
\item First, it is expected that the system will decay into and out of a plateau value which is equal to the critical value of the nonergodicity factor. Therefore, the optimal domain will yield $f_a\approx f_b \approx f_q^*$.
\item Second, for the early part of the $\beta$-relaxation decay, note that $\tau_0$ is a constant independent of both wavenumber and density. Therefore, the optimal domain over this part of the data can be found and set once; it will yield identical results as we change wavenumber and as we change density (provided we remain ``close" to the transition where $\epsilon$ is small).
\item Third, the time scale $\tau_{\alpha}$ depends on density, but not on wavenumber. The optimal domain over the late part of the $\beta$-regime will yield consistent values for $\tau_{\alpha}$ as we vary $q$.
\item Finally, the optimal time scale will show the smallest residual between data and fit. Visually, we will verify this by plotting $\log_{10}(|F(q,t)-f_q^*|)$ versus $\log_{10}t$ where the fit is expected to be a straight line and where deviations from the data will be most clear.
\end{itemize}

In Figs. \ref{fig:Ffixed1} and \ref{fig:Ffixed2}, we show our the results on either side of the transition for $\eta = 0.601497$ and $\eta = 0.601498$ at $q=7.45$ near the static structure factor maximum. As can be seen visually, each domain is only a few decades long, but the data match the model well and the exponent values capture the decays appropriately.

\begin{figure}[p]
\includegraphics[width=\columnwidth]{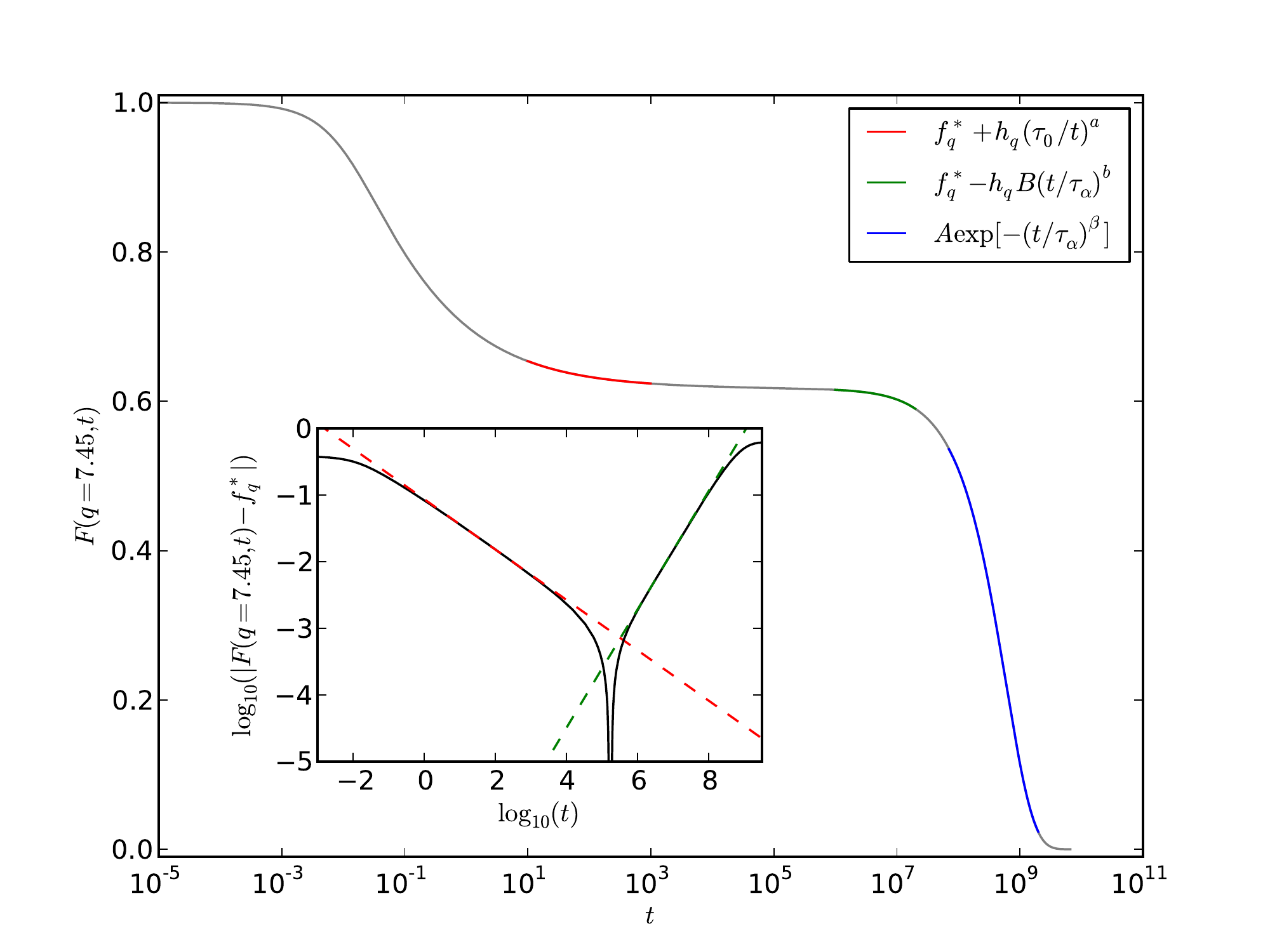}
\caption{Values of $a$ and $b$ were found as the solution to the set of coupled equations for $\lambda$ and $\gamma$ where the value of $\gamma$ was found by the fit of $\tau_q$ as a function of density. These values were used along with the value of $h_q$ found from the nonergodicity factor fit as fixed parameters in fits to the early and late portions of the $\beta$-relaxation of $F(q,t)$. Shown here is one such fit for $\eta = 0.601497$ just below the transition and $q=7.45$ near the first structure factor maximum. In addition to the two power-law fits, we include the stretched exponential fit of the $\alpha$-regime. The inset shows the power law fits plotted on a log-log scale. Several decades of data lie along the fit functions showing good agreement with the model.}
\label{fig:Ffixed1}
\end{figure}

\begin{figure}[p]
\includegraphics[width=\columnwidth]{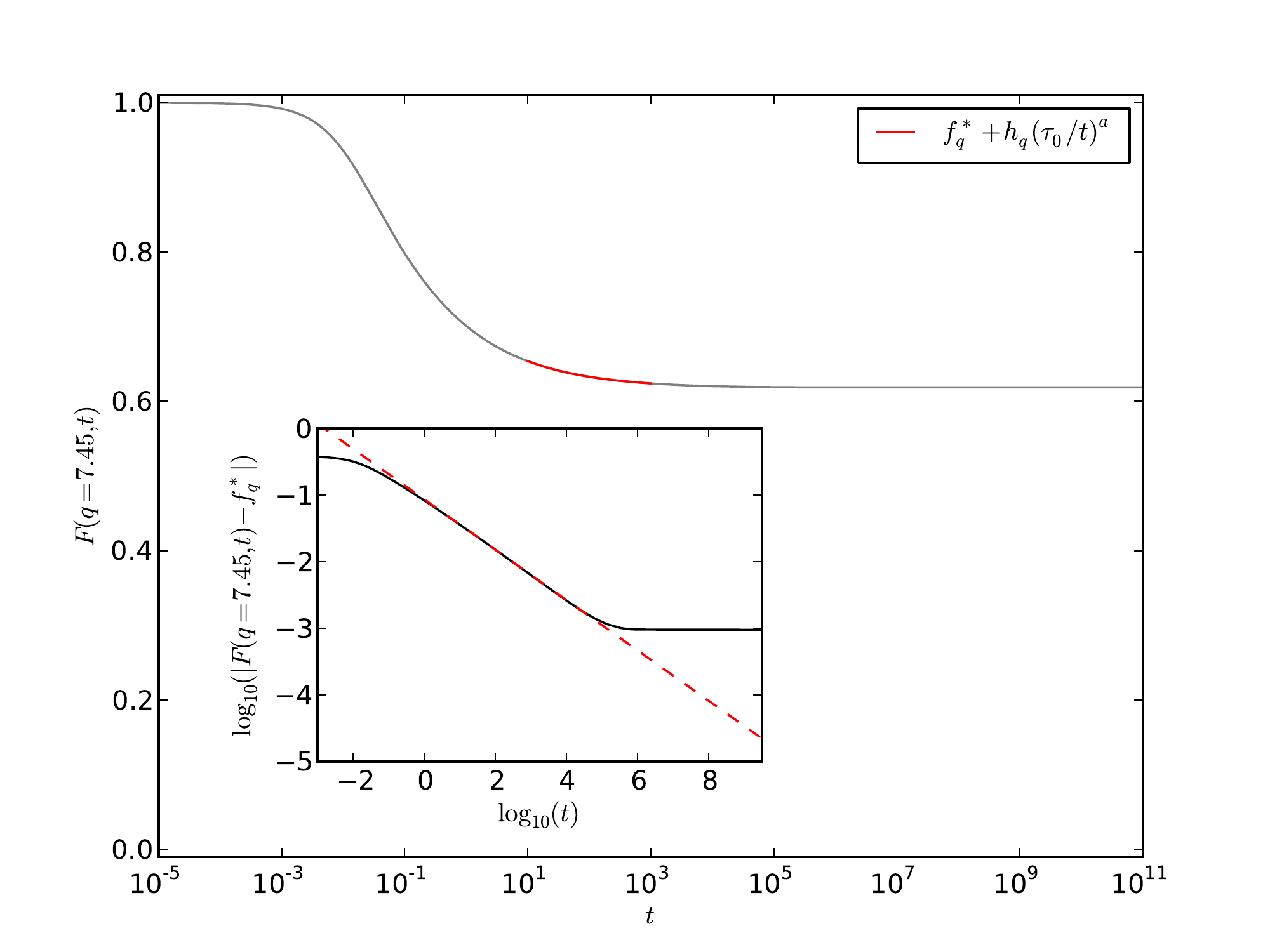}
\caption{Values of $a$ and $b$ were found as the solution to the set of coupled equations for $\lambda$ and $\gamma$ where the value of $\gamma$ was found by the fit of $\tau_q$ as a function of density. These values were used along with the value of $h_q$ found from the nonergodicity factor fit as fixed parameters in fits to the early and late portions of the $\beta$-regime of $F(q,t)$. Shown here is the single power-law fit for $\eta = 0.601498$ just above the transition at $q = 7.45$. The inset shows the power law fit plotted on a log-log scale. We see that several decades of data lie along the fit function showing good agreement with the model.}
\label{fig:Ffixed2}
\end{figure}

\subsubsection{Direct fit to the intermediate structure factor, F(q,t)}
While the above verification worked well, it it valid to ask if the exponent values (and other fit values) could be found directly from the $\beta$-regime data. As cautioned in the introduction to this section, this is less straightforward than expected due to the variability caused by domain choice, but we will show here that if care is taken, it can be done and the results are consistent with what we found above.

Again, let us perform a constrained fit on the function $F(q,t)$ in the ergodic phase using Eqs.~(\ref{eq:fit1}) and (\ref{eq:fit2}), but this time allowing $a$ and $b$ to vary. (The critical amplitude $h_q$ remains fixed.) In what follows, we will consider two cases. In the first, which we call the unconstrained case, $a$ and $b$ are allowed to take the best fit value from minimizing the squares of the residuals. In the second, which we call the constrained case, $a$ and $b$ must additionally satisfy the exponent constraint, Eq.~({\ref{eq:lambda}).

In addition to the criteria setting the optimal domain outlined above, we may add the following:
\begin{itemize}
\item The optimal domain will be the one where the constrained and unconstrained fits yield identical choices for $a$ and $b$.
\end{itemize}

Doing this fit for values of density just below the transition, we find robust values for the exponents
\begin{equation}
\begin{array}{ccc}
a=0.374(10),~ & b=0.878(8),~ & \lambda=0.568(16).
\end{array}
\end{equation}
Uncertainties here again represent a small spread in values from fits at different wavenumber and density and are slightly larger than in the previous method. The results, however, are consistent and we see that a suitable and reliable method is possible despite the difficulties of the fit.

\section{Comparison with other work}
\label{sec:comparison}
If one wishes to compare our work here on hard spheres obeying Smoluchowski dynamics to experiment, the closest realization is a colloidal suspension\cite{pusey86}.

Macroscopic particles (typically on the order of 100 nm - 1000 nm) are suspended in a mediating fluid leading to Brownian motion. Particles interact with each other only through repulsion on contact---a hard sphere potential, to good approximation---and are large enough to be observed by confocal microscopy so individual particle motions can be tracked. The most widely-used colloid for these hard sphere-type interactions is poly-methyl-methacrylate (PMMA), which can be density-matched to the liquid (eliminating the effect of gravity and the consequence of sedimentation) and used with charge screening solvents (eliminating any van der Waals and electrostatic-type effects which would otherwise give the potential a ``soft" attractive component). A comprehensive survey of colloidal experiments is available in Ref. \onlinecite{weeksbook}.

While our theory (as well as MCT) has been developed under the assumption of monodisperse particles of identical size, such systems crystalize easily and at lower densities than the predicted ergodic-nonergodic transition; they therefore cannot be used to investigate the glass transition. Instead, experiments typically introduce some polydispersity (approximately 10\%) and allow the particle sizes to vary about the average. This spread in values prevents crystallization and methods are available for determining an effective single particle packing fraction for comparison with monodisperse systems and theories. When creating such suspensions it is straightforward to determine relative differences in packing fractions (typically as small as $10^{-4}$), but there is some difficulty in assigning an absolute value to the packing fraction; uncertainties can be as much as a few percent and have led to difficulty comparing different experiments\cite{weeksbook}.

Beginning soon after after mode-coupling theory was developed, many colloidal PMMA experiments have been performed to check the validity of the theory\cite{elmasri,pusey87,vanmegen93, vanmegen94,vanmegen98}. This work established that the MCT prediction for the glass-transition density of $\eta^* = 0.51(1)$\cite{barrat} is too low, though measurements of the actual transition density have been in conflict; typical values span a range $0.57 \leq \eta^* \leq 0.60$, up against the limit of what can be made and measured in the lab. Results, however, do lend credence to the predicted power-law forms describing the $\beta$-relaxation, and the MCT exponents---$\lambda = 0.734$, $a=0.312$, and $b=0.584$\cite{spyridis}---are plausible given the limits of power-law fits to sparse data\cite{goetze99}.

Advances in recent years have allowed for both more sophisticated experiments and simulations. In the most complete colloidal PMMA study to date, dynamic light scattering is used to measure concentrations of colloidal hard spheres up to a density of $\eta=0.5970$\cite{brambilla}. Though the system remains ergodic over the entire range studied, there is tremendous slowing down and over seven decades of relaxation time, $\tau_{\alpha}$. The authors find that a fit of these relaxation times for $0.49 < \eta \leq 0.585$ yields a power-law divergence (Eq.~(\ref{eq:taualpha})) at critical density $\eta^* = 0.590(5)$ with power-law exponent $\gamma = 2.5(1)$.

This result, however, contradicts the experiment; the system clearly remains ergodic above this critical density. To resolve this, the authors find that when including data for $\eta > 0.585$, the divergence is better modeled as an exponent of the Vogel-Fulcher-Tammann form\cite{vogel,fulcher,tammann}
\be
\tau_{\alpha}(\eta) = \tau_{\infty}\exp\bigg[\frac{A}{(\eta_0-\eta)^{\delta}}\bigg]
\label{eq:VFT}
\ee
with $\delta = 2.0(2)$ and $\eta_0 = 0.637(2)$. The authors interpret the avoidance of the power-law type divergenceas evidence of a crossover to an ``activated" regime where hopping modes not addressed by MCT (or our theory) become important. This interpretation is debated\footnote{In follow-up comments, the authors defend this interpretation against claims from two other groups \cite{brambillacomments1, brambillacomments3}. The challengers argue -- by slightly different methods -- that uncertainties in the absolute density or in the polydispersity of the colloid radii lead to enough ambiguity  that the data is in fact consistent with a power-law divergence over the full range, putting the transition at slightly higher density, $\eta^* \sim 0.595-0.6$; under this interpretation, there would be no need to claim a crossover to an {``activated"} range. The original authors, however, argue that their control over particle size is within reason and present counter-arguments to each claim \cite{brambillacomments2, brambillacomments4}. The debate over the existence (or even possibility) of a crossover in colloidal particles remains active; see the following footnote for more.} and it is unclear if such hopping processes should even be accessible in a hard sphere or colloidal system\footnote{There remains substantial debate as to whether or not a hard-sphere or colloidal system avoids (or is theoretically able to avoid) the power law divergence predicted by mode-coupliing theory (and our own theory, Eq.~(\ref{eq:taualpha})) in favor of the activated exponential divergence (Eq.~(\ref{eq:VFT})) seen in the Brambilla work\cite{brambilla}. Early extentions to MCT suggests that MCT is virtually exact for colloids and the power-law should hold\cite{dasmazenko}, but others point to free-volume arguments that predict the exponential divergence with $\delta = 1$\cite{cohen}. Simulation work pre-dating Brambilla \textit{et al.} supports exponential divergence, (possibly with $\delta > 1$)\cite{szamel04,berthier07}, as does some new theoretical explorations of MCT\cite{cates,ABL}, but reasons that the power-law form should be avoided remain controversial. While there is work showing exponential divergence for granular materials approaching the limit of random close packing \cite{torquatto,ohern,schweizer}, there is no concensus that this dynamic arrest is equivalent to the glass-transition\cite{witten}.}. This work, however, remains the only data set with sufficient resolution at such high densities to allow comparison between the two divergence forms at all.

In the same study, the authors also run Monte Carlo hard sphere simulations. Power-law divergence fits to the same density ranges above yielded an identical critical density $\eta^* = 0.590(5)$ and critical exponent $\gamma = 2.5(1)$. Fitting to the alternate exponential-divergence yielded an even higher divergence density $\eta_0 = 0.651(2)$ than the experiment.

Though this work covers a wide time range, the authors make no attempt to fit power-law functions to the decays into and out of the plateau of the $\beta$-regime. However, coupling the authors' value for $\gamma$ with the $\lambda$-constraint (just as we do for our own work here), the predicted exponents are $a=0.308(10)$ and $b=0.57(3)$, with $\lambda = 0.743(20)$. These are consistent with the prediction of MCT using Percus-Yevick hard spheres quoted above, but different from the values we find in our own solution.

\section{Conclusion}
\label{sec:conclusion}
We have reviewed a new theory which derives the governing kinetic and static equations for $n$-point cumulant functions between density and response fields for systems of particles obeying either Newtonian or Smoluchowski dynamics in a self-consistent perturbation expansion in the effective interparticle potential. At second order, we find that the density-density cumulant obeys a kinetic equation similar to that seen in mode-coupling theory with a memory function quadratic in the potential and quadratic in the cumulant itself. In the longtime, static limit, this equation supports an ergodic-nonergodic transition and we outline an asymptotic analysis that predicts a two-step decay with associated diverging time scales, and power-law decays into and out of an extended plateau with wavenumber- and density-independent exponents.

For a simple model system of hard spheres obeying Smoluchowski dynamics, we solve for the full behavior of the intermediate structure factor at all wavenumbers and over a wide range of densities from dilute fluid to maximally dense. Near the transition, the intermediate structure factor evolves from an exponential decay into the two-step decay characteristic of supercooled liquids, and the relaxation slows by orders of magnitude. At even higher density, we pass the critical packing fraction and the correlation function decays only to a finite plateau, showing that the system has naturally selected the nonergodic solution. This work represents the first numerical solution showing the full dynamics near an ergodic-nonergodic transition outside of mode-coupling theory and covers an unprecedented nearly 15 decades of scaled time. The longtime results of the full dynamics match earlier studies of the statics under the same theory and we recover the previously investigated nonergodicity factors and critical transition density $\eta^* = 0.60149761(10)$.

As part of this solution, we test the results using two forms for the memory function at second-order in the potential. The first is the so-called longtime form where were drop the self-contribution and simplify the collective contribution. The second is a a more complete form which keeps all terms, but uses the zeroth-order approximation for the vertex functions. In this latter solution, we again monitor the density-density time correlation function $F(q,t)$, but also compute two dressed propagators, $\bar{F}(q,t)$ and $\tilde{F}(q,t)$, which modify short-time behavior, but not the long-time limit. We find that the two solutions behave exactly as expected and that the universal features of the approach to the ergodic-nonergodic transition are identical in both cases. In fact, the two solutions can be mapped onto each other with a simple scaling of the microscopic time $\tau_0\rightarrow\tau_0^{\prime}$ in all the appropriate equations. This equivalence justifies \textit{a postori} the assumptions implicit in the asymptotic expansion approximation and reinforces the similarities seen between the MCT memory function and the longtime approximate memory function derived here.

Using the full dynamic solution, we were able to investigate the two time scales which are relevant in the ergodic regime---$\tau_{\alpha}$ which sets the time scale of the longtime relaxation to zero and $\tau_{\beta}$ which sets the scale of the time spent in the intermediate plateau. Both time scales are seen to diverge as the density increases to the critical density and from the power-law scaling of $\tau_{\alpha}$, we extract our first critical exponent, $\gamma = 1.887(4)$. This exponent is in turn related to the other two exponents as $\gamma = 1/2a + 1/2b$. In conjunction with the critical exponent parameter,
\be
\lambda = \frac{\Gamma^2(1-a)}{\Gamma(1-2a)} = \frac{\Gamma^2(1+b)}{\Gamma(1+2b)},
\ee
we found $a = 0.375(3)$, $b=0.887(4)$, and $\lambda=0.5587(18)$. These are consistent with parameters extracted from direct fits to $F(q,t)$ where $a = 0.374(10)$, $b=0.878(8)$, and $\lambda=0.568(16)$.

In functional form, our solution matches the most important features seen in experiment and simulation and predicted by mode-coupling theory. The transition density we find at this order is compatible with the range of values seen in polydisperse colloidal suspensions and associated hard sphere simulations, but our universal exponents do differ from these and from MCT. Additionally, our nonergodicity factor $f_q$ and critical amplitude $h_q$ share qualitative features with measured/predicted forms, but do not match quantitatively. Crucially, however, this need not be the end of the story. Unlike MCT, this theory has a clear method for calculating corrections and we are in a position to explore how the physics evolve order-by-order. Understanding how the values of these universal exponents and form factors change as we include higher-order terms will be an interesting future undertaking.

The numerical solution here establishes this new theory as a viable alternative to mode-coupling theory and one derived from first principles with a self-consistent expansion that is well-motivated on physical grounds. Many salient features of the glass-transition emerge naturally and we see that the results can be analyzed with the same critical dynamics machinery pioneered in MCT.

The first parameter estimates here are tantalizing, but a host of new tests await. This theory can be extended to study multi-component systems or systems trapped in external fields\cite{henderson, zaccarelli}; to look at four-point correlation functions and investigate the growing length scales thought to be associated with dynamic heterogeneity\cite{berthier04, berthier05, BB, BBMR, garrahan}; or to further investigate the equivalence between the longtime limits of Newtonian and Smoluchowski dynamics\cite{ULTD,szamel04}.

\begin{acknowledgments}
The author would like to extend his sincerest thanks to Professor Gene Mazenko for his guidance and direction on this work and for guiding him through many years as a student. The author would also like to thank Paul Spyridis for many helpful conversations and specifically to acknowledge Mr. Spyridis's help improving numerical algorithms for solving the kinetic equation quickly and efficiently. This work was supported by the Department of Physics at the University of Chicago and by the United States Department of Education under a Graduate Assistance in Areas of National Need (GAANN) fellowship.
\end{acknowledgments}

\newpage
\appendix
\section{Method of numerical solution}
\label{app:numerical}
Numerically solving the integro-differential kinetic equation, Eq.~({\ref{eq:kinetic}), is a difficult task. As discussed in the introduction, the density-density time correlation function decays according to several very different time scales as one changes density  from dilute fluid to dense fluid to glass, and as one progresses in time from microscopic dynamics through the $\beta$-relaxation and finally to the $\alpha$-relaxation. Additionally, the behavior at late times depends non-trivially on the behavior at short times through the convolution integral over the memory function.

It is therefore important to develop a method which treats the quick decay at short times and small densities differently than the long decays of late times and large densities. Our approach builds up a solution by starting at the initial conditions and advancing one time step at a time. Rather than keeping a constant time step, though, we introduce a variable time step size and rewrite the kinetic equation and the memory function in forms which separate late- and early-time quantities. Such rewrites let us introduce memory-saving tricks while maintaining sufficient accuracy, which in turn allows us to extend our solution out dozens of decades of time and explore systems very close to the transition point.

\subsection{Casting the kinetic equation in a tractable form}
Our approach will be to begin with the $t=0$ initial conditions---$F(q,t=0) = 1$ and $\partial F(q,t)/\partial t|_{t=0} = 0$---and advance our solution forward in small time steps, updating each important quantity as we go. Functions like $F(q,t)$ and $K(q,t)$ will be stored in arrays at discrete $q$ and $t$ values, but such arrays grow larger and consume more memory with each time step. Proceeding na\"{i}vely with a constant step size means that each new calculation takes longer than the last; integrals over time require summing more and more terms eating up more and more memory and computation time. Computational limits are reached before the solution can be extended to sufficiently late times.

Therefore, we want to rewrite our equations so we can introduce a variable step size while also maintaining sufficient accuracy. This means separating short-time and long-time quantities and structuring our approximate integrals so that we are storing only small amounts of data to memory which are constantly updated rather than appending onto existing arrays.

Let us begin with the integral term of the kinetic equation. First, split the integral into two time pieces at $s=t/2$ and integrate the first by parts. We find, after rearrangement,
\be
\int_0^{t} ds K(q,t-s)\frac{\partial}{\partial s} F(q,s)
&=&K(q,t-t/2)F(q,t/2)-K(q,t)F(q,0)
\nonumber\\
&&+\int_0^{t/2} ds F(q,s)\frac{\partial}{\partial t} K(q,t-s)
\nonumber\\
&&+\int_0^{t-t/2} ds K(q,s)\frac{\partial}{\partial t} F(q,t-s).
\ee
Notice that the integral is now expressed so that the \emph{functions} are evaluated at ``early" times ($0 \leq s \leq t/2$) and the \emph{derivatives} are evaluated at ``late" times ($t/2 \leq t-s \leq t$).

Discretizing time with step $\Delta_t$ and wavenumber step $\Delta_q$ such that
\be
t\rightarrow t_j \equiv j\Delta_t,~~~ 0\leq j < N_t
\ee
and
\be
q\rightarrow q_i \equiv i\Delta_q,~~~ 0\leq i < N_q,
\ee
we may introduce the notation for generic function
\be
g(q_i,t_j) \rightarrow g[i][j].
\ee

Let us next break up the integrals into sums over smaller integrals. For example,
\be
\int_0^{t/2} ds F(q,s)\frac{\partial}{\partial t} K(q,t-s)
\approx \sum_{m=1}^{j/2} \int^{t_m}_{t_{m-1}} ds F(q,s)\frac{\partial}{\partial t} K(q,t-s).
\ee
Because the derivatives are evaluated at ``late" times where they change slowly relative to the step size, they are nearly constant and we can take each outside its integral, approximating as
\be
\int_0^{t/2} ds F(q,s)\frac{\partial}{\partial t} K(q,t-s)
\approx \sum_{m=1}^{j/2} \bigg[\frac{K(q,t_j-j_{m-1})-K(q,t_j-t_m)}{\Delta_t}\bigg]\int^{t_m}_{t_{m-1}} ds F(q,s).
\nonumber\\
\ee
Approximating the remaining integral using the midpoint method, we define
\be
dF(q,t_m)\equiv \frac{1}{\Delta_t}\int^{t_m}_{t_{m-1}} ds F(q,s) \approx \frac{F(q,t_m) +F(q,t_{m-1})}{2}.
\ee
and
\be
dK(q,t_m)\equiv \frac{1}{\Delta_t}\int^{t_m}_{t_{m-1}} ds K(q,s) \approx \frac{K(q,t_m) +K(q,t_{m-1})}{2}.
\ee
such that
\be
\int_0^t ds K(q,t-s)\frac{\partial}{\partial s} F(q,s)
&\approx&K(q,t_{j-j/2})F(q,t_{j/2})-K(q,t_j)F(q,0)
\nonumber\\
&&+\sum_{m=1}^{j/2} \bigg[K(q,t_j-t_{m-1})-K(q,t_j-t_m)\bigg]dF(q,t_m)
\nonumber\\
&&+\sum_{m=1}^{j-j/2} \bigg[F(q,t_j-t_{m-1})-F(q,t_j-t_m)\bigg]dK(q,t_m).
\ee

Next, let us approximate the derivative term of the kinetic equation via a third-order, backward form,
\be
\frac{\partial}{\partial t} F(q_i,t_j) = \frac{(3/2)F[i][j]-2F[i][j-1]+(1/2)F[i][j-2]}{\Delta_t}.
\ee
After much wrangling, the kinetic equation can be rearranged as
\be
F[i][j] = A[i]*K[i][j] + B[i][j]
\ee
where
\be
A[i] = \frac{S[i](F[i][0]-dF[i][1])}{1 + S[i] dK[i][1]+3S[i]/(2(i\Delta_q)^2\Delta_t)}
\ee
and
\be
B[i][j] &=& \bigg\{\frac{2S[i]}{(i\Delta_q)^2\Delta_t}F[i][j-1]
-\frac{S[i]}{2(i\Delta_q)^2\Delta_t}F[i][j-2]
-S[i]\bigg[ K[i][j-j/2]F[i][j/2]
\nonumber\\
&&-K[i][j-1]dF[i][1] + \sum_{s=2}^{j/2} \bigg(K[i][j-(s-1)]-K[i][j-s]\bigg)dF[i][j]
\nonumber\\
&&-F[i][j-1]dK[i][1] + \sum_{s=2}^{j-j/2} \bigg(F[i][j-(s-1)]-F[i][j-s]\bigg)dK[i][j]\bigg]\bigg\}
\nonumber\\
&&/\bigg[1 + S[i] dK[i][1]+3S[i]/(2(i\Delta_q)^2\Delta_t)\bigg].
\ee

Note that neither $A[i]$ nor $B[i][j]$ depends on $F[i][j]$ or $K[i][j]$. And, note that when $q=0$, $A[i]=0$ and $B[i][j] = 1$ leading to $F[0][j] = 1$ as expected.

\subsection{Solving the Equation}
Now that we have a discrete version of the kinetic equation, we can numerically approximate $F(q,t)$ for a given memory function $K(q,t)$. (We will address $K(q,t)$ explicitly in the next section.) We want to implement a step-doubling routine so that we can have fine spacing at small times (for accuracy) and larger spacing at late times (for speed).

The algorithm is as follows:

\begin{enumerate}[A.]
\item
Choose a small initial time step $\Delta_t = 1\times 10^{-7}$. For all values of $q$ and for values of $0 \leq t < \Delta_t N_t/2$, initialize the $F[i][j]$ array using the approximate short-time solution $F^{(0)}(q,t)=e^{-q^2t/S(q)}$. Compute the corresponding $K[i][j]$, $dF[i][j]$ and $dK[i][j]$.
\item
For each value of $\Delta_tN_t/2 \leq t < \Delta_tN_t$, complete a short iteration scheme of the following form:
\begin{enumerate}[1.]
\item
Compute $A[i]$ and $B[i][j]$;
\item
Make the initial guess $F[i][j] = F[i][j-1]$ for all $i$;
\item
Compute $K[i][j]$ for all $i$;
\item
Update $F[i][j]$ via $F[i][j] = A[i]*K[i][j]+B$ for all $i$; and
\item
Repeat steps 2 through 4 until $F[i][j]$ converges. (Convergence is relatively quick.)
\end{enumerate}
Compute the corresponding $dF[i][j]$ and $dK[i][j]$.
\item
Perform a step doubling $\Delta_t\rightarrow 2\Delta_t$ and map the functions as
\be
F[i][j] &=& F[i][2j]
\nonumber\\
K[i][j] &=& K[i][2j]
\nonumber
\ee
and the derivatives (by Simpson's rule over the new larger interval) as
\be
dF[i][j] = (F[i][2j] + 4F[i][2j - 1] + F[i][2j - 2])/6
\nonumber\\
dK[i][j] = (K[i][2j] + 4K[i][2j - 1] + K[i][2j - 2])/6.
\nonumber
\ee
\item
Repeat B and C until $F(q,t)$ has been solved to sufficiently large $t$.
\end{enumerate}

This method keeps the partial integrals up-to-date at high accuracy (but low memory usage) while allowing derivatives to be calculated on the fly.

After much testing, we have chosen optimal program parameters.

For time discretization, we use an initial time step $\Delta_t = 1\times 10^{-7}$ with number of time steps $N_t = 200$. This gives us a ``jumpstart" (the short-time approximation described in item A of the algorithm above) up to $t=\Delta_t N_t/2 = 1\times 10^{-5}$. The program is fairly robust to these choices as long as the ``jumpstart" does not extend to time beyond $t\approx5\times 10^{-5}$. This is consistent with our findings that the microscopic time $\tau_0$ (where the behavior is liquid-like regardless of density) is of order $10^{-3}$.

For wavenumber discretization, we use wavenumber step $\Delta_q = 0.05$ with number of wavenumber steps $N_q = 2000$. This leads to an integral cutoff at $q_{max} = \Delta_q N_q = 100$. As long as $q_{max} > 50$, we find no dependence in the solution on $N_q$.

\subsection{Simplifying the memory function}
The memory function is, in general, a 3-dimensional integral. The azimuthal integration is trivial, but that leaves two nested integrals which can be time intensive to compute. It is, however, possible to decouple this double integral into to two single integrals and save significant computation time.

We begin by rewriting the (longtime approximate) memory function as
\be
K^{(LT)}(q,t) &=& \frac{\pi}{12\eta}\int \frac{dk d\theta~2\pi k^2\cos\theta}{(2\pi)^3} V(k) S(k) F(k,t)
V(|{\bf q}-{\bf k}|) S(|{\bf q}-{\bf k}|) F(|{\bf q}-{\bf k}|,t)
\nonumber\\
&=&\frac{1}{48\pi\eta}\int dk du ~k^2 V(k) S(k) F(k,t)
\nonumber\\
&&\times V(\sqrt{q^2+k^2+2qku}) S(\sqrt{q^2+k^2+2qku}) F(\sqrt{q^2+k^2+2qku},t)
\nonumber\\
&=& \frac{1}{48\pi\eta q}\int_0^{\infty} kdk V(k) S(k) F(k,t) \int_{|q-k|}^{q+k} pdp V(p) S(p) F(p,t)
\ee
where we've made the substitution $u = \sin\theta$ in the second line and $p^2 = q^2+k^2+2qku$ in the third.

Let us define the following:
\be
\sigma(x) \equiv \int_0^x pdp V(p)S(p)F(p,t).
\ee
Using this notation, we can write our memory function as
\be
K^{(LT)}(q,t) &=& \frac{1}{48\pi\eta q}\int_0^{\infty} kdk V(k) S(k) F(k,t) \bigg(\sigma(q+k)-\sigma(|q-k|)\bigg).
\ee

Using our discritization scheme from above, we may approximate the memory function by the Euler method as
\be
K^{(LT)}(q,t) &=& \frac{1}{48\pi\eta q}\int_0^{\infty} kdk V(k) S(k) F(k,t) \bigg(\sigma(q+k)-\sigma(|q-k|)\bigg)
\nonumber\\
&\approx&\frac{1}{48\pi\eta (i\Delta_q)}
\sum_{n=0}^{N_q-1} (n\Delta_q)\Delta_q V[n]S[n]F[n][j] \bigg(\sigma[i+n] -\sigma[|i-n|]\bigg)
\ee
where
\be
\sigma[x] &=& \sum_{n=0}^{x} (n\Delta_q)\Delta_q V[n]S[n]F[n][j]
\nonumber\\
&=& \sigma[x-1] + (n\Delta_q)\Delta_q V[n]S[n]F[n][j].
\ee

\section{The dressed propagators and the no-vertex approximation solution}
\label{app:dressed}
The dressed propagators which naturally fall out of the theory initially seem to be a great complication, but can be treated with the same techniques developed in Appendix \ref{app:numerical}. Solving the kinetic equation using the full second-order memory function without vertex corrections, Eq.~(\ref{eq:NoV}), shows us that $\bar{F}(q,t)$ and $\tilde{F}(q,t)$ behave as predicted---they decay quicker than $F(q,t)$ at short times, but approach $F(q,t)$ at long times. Plots of all three correlations functions are given in Fig. \ref{fig:dressed}.

\begin{figure}[p]
\includegraphics[width=\columnwidth]{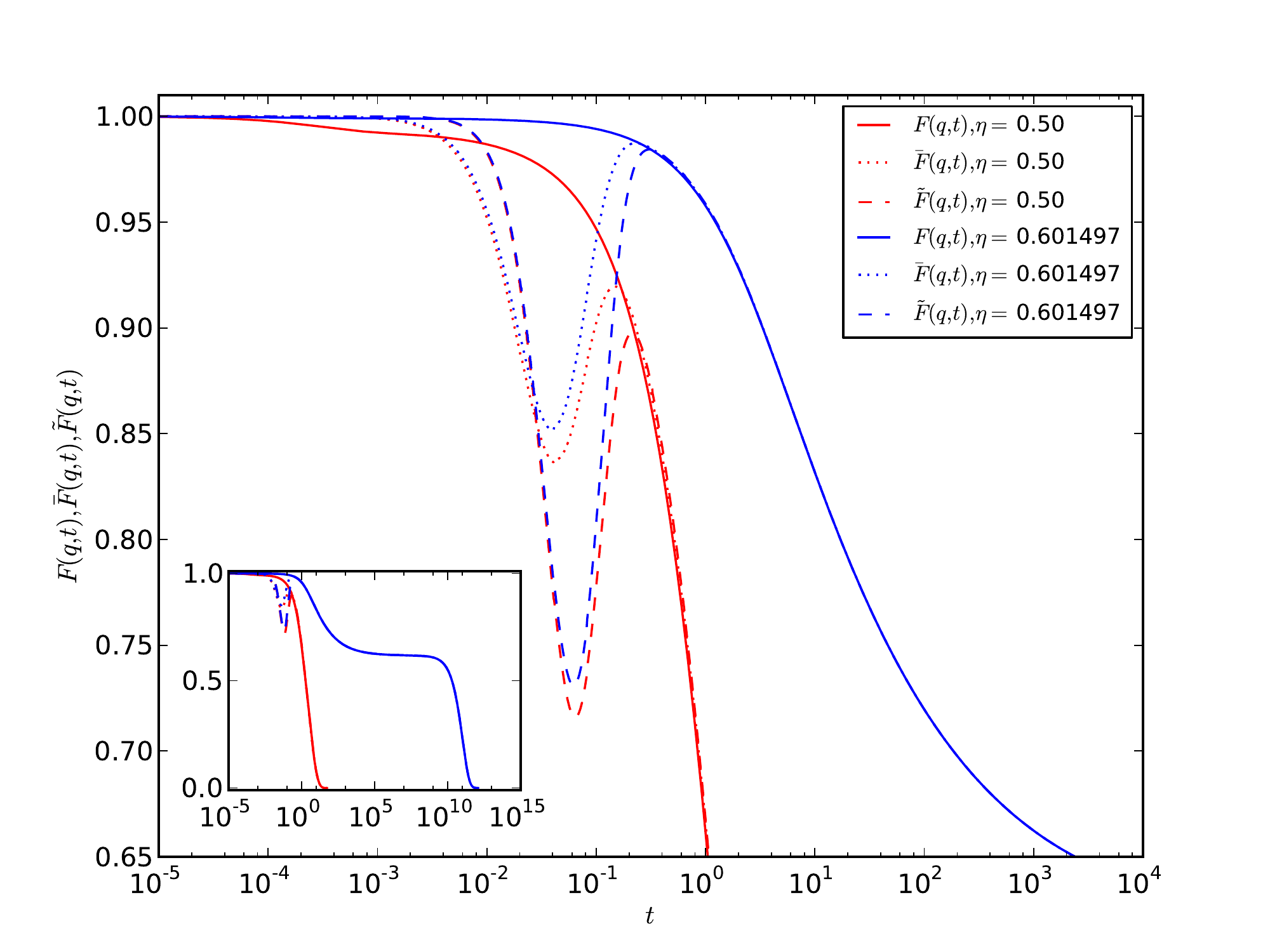}
\caption{The dressed propagators $\bar{F}(q,t)$ and $\tilde{F}(q,t)$ are shown alongside the full intermediate static structure factor $F(q,t)$ at wavenumber $q=7.45$ and at two densities---$\eta = 0.5$ well below the transition and $\eta = 0.601497$ very close to the transition. The full functions are plotted in the inset, but the main plot shows just the shorttime portion where the functions differ. Note that the dressed propagators decay quickly, but then changes sign and actually increases to meet $F(q,t)$. Despite the very different longtime behaviors, the dressed propagators at these two packing fractions both ``dip" to roughly the same values at roughly the same times meaning that this anomalous behavior is not highly dependent on density.}
\label{fig:dressed}
\end{figure}

Furthermore, the solution for the full function $F(q,t)$ is identical to that found with the longtime approximate form for the memory function, Eq.~(\ref{eq:LT}), with a rescaling of the microscopic time $\tau_0\rightarrow\tau_0^{\prime}$. Both solutions show the same two-step decay with identical exponents, and the time scales $\tau_{\alpha}$ and $\tau_{\beta}$ show the same power-law divergences. The nonergodicity factors and the critical density at which the ergodic-nonergodic transition occurs are likewise identical.

In the following sections, we include for completeness the discretizations and approximations used to compute the dressed propagators and solve the kinetic equation for the no-vertex corrections memory function form.

\subsection{Approximating $\bar{F}(q,t)$}
The first dressed propagator is given by
\be
\bar{F}(q,t) = F^{(0)}(q,t) +q^2\int_0^t dsF(q,s)F^{(0)}(q,t-s)
\label{eq:Fbar}
\ee
which can be split and integrated by parts to give
\be
\bar{F}(q,t) &=& F^{(0)}(q,t) -F^{(0)}(q,t-t/2)F(q,t/2)+F^{(0)}(q,0)F(q,t)
\nonumber\\
&&+q^2\int_0^{t/2} ds F(q,s)F^{(0)}(q,t-s)+\int_0^{t-t/2} ds F^{(0)}(q,s)\frac{\partial}{\partial s}F(q,t-s).
\ee

The first integral can be approximated as
\be
q^2\int_0^{t/2} ds F(q,s)F^{(0)}(q,t-s)
&\approx& q^2 F^{(0)}(q,t) \sum_{m=1}^{j/2}\int_{t_{m-1}}^{t_{m}} ds F(q,s)e^{q^2s},
\ee
and, if we define
\be
d\bar{F}(q_i,t_m) \equiv \frac{1}{\Delta_t}\int_{t_{m-1}}^{t_{m}} ds F(q_i,s)e^{q_i^2s}
\approx\frac{e^{q_i^2 t_m}F(q_i,t_m)+e^{q_i^2 t_{m-1}}F(q_i,t_{m-1})}{2},
\ee
we then have
\be
q^2\int_0^{t/2} ds F(q,s)F^{(0)}(q,t-s)
=q^2\Delta_tF^{(0)}(q,t)\sum_{m=1}^{j/2}d\bar{F}(q,t_m).
\ee
The second integral similarly becomes
\be
\int_0^{t-t/2} ds F^{(0)}(q,s)\frac{\partial}{\partial s}F(q,t-s)
&\approx&\sum_{m=1}^{j-j/2}\bigg[\frac{F(q,t-t_m)-F(q,t-t_{m-1})}{\Delta_t}\bigg]
\nonumber\\
&&\times \bigg(\frac{-1}{q^2}\bigg)[e^{-q^2t_m}-e^{-q^2t_{m-1}}].
\ee

Plugging these terms back into Eq.~(\ref{eq:Fbar}), we compactly write
\be
\bar{F}[i][j] = C[i]F[i][j]+D[i][j]
\ee
where
\be
C[i] = \frac{F^{(0)}[i][0]\bigg[(i\Delta_t)^2\Delta_t-1\bigg]+F^{(0)}[i][1]}{(i\Delta_q)^2\Delta_t}
\ee
and
\be
D[i][j] &=& F^{(0)}[i][j]\bigg[1+(i\Delta_q)^2\Delta_t\sum_{m=1}^{j/2}d\bar{F}[i][m]\bigg]-F^{(0)}[i][j-j/2]F[i][j/2]
\nonumber\\
&&+\bigg(\frac{1}{(i\Delta_q)^2\Delta_t}\bigg)\sum_{m=2}^{j-j/2}\bigg[F[i][j-m]-F[i][j-m-1]\bigg]
\bigg[F^{(0)}[i][m-1]-F^{(0)}[i][m]\bigg]
\nonumber\\
&&+F[i][j-1]\bigg[\frac{F^{(0)}[i][0]-F^{(0)}[i][1]}{(i\Delta_q)^2\Delta_t}\bigg].
\ee

At $q=0$, these reduce to $C[0] = 1$ and $D[0] = 0$ yielding $\bar{F}[0][j] = 1$
as expected.

\subsection{Approximating $\tilde{F}(q,t)$}
The second dressed propagator is given by
\be
\tilde{F}(q,t) = F^{(0)}(q,t)[1+q^2t] +q^4\int_0^t ds(t-s)F(q,s)F^{(0)}(q,t-s)
\label{eq:Ftilde}
\ee
which can be rewritten as
\be
\tilde{F} &=& F^{(0)}(q,t)[1+q^2t] -F^{(0)}(q,t-t/2)F(q,t/2)[1+q^2(t-t/2)] +F^{(0)}(q,0)F(q,t)
\nonumber\\
&&+q^4\int_0^{t/2} ds (t-s)F(q,s)F^{(0)}(q,t-s)
+\int_0^{t-t/2} ds (1+q^2s)F^{(0)}(q,s)\frac{\partial}{\partial s}F(q,t-s).
\nonumber\\
\ee

The first integral can be simplified as
\be
&&q^4\int_0^{t/2} ds (t-s)F(q,s)F^{(0)}(q,t-s)
\nonumber\\
&\approx& q^4 F^{(0)}(q,t)\bigg[t\sum_{m=1}^{j/2}\int_{t_{m-1}}^{t_{m}} ds F(q,s)e^{q^2s}
-\sum_{m=1}^{j/2}\int_{t_{m-1}}^{t_{m}} ds s F(q,s)e^{q^2s}\bigg],
\ee
and, if we define
\be
d\tilde{F}(q_i,t_m) \equiv \frac{1}{\Delta_t}\int_{t_{m-1}}^{t_{m}} ds sF(q_i,s)e^{q_i^2s}
\approx\frac{t_me^{q_i^2 t_m}F(q_i,t_m)+t_{m-1}e^{q_i^2 t_{m-1}}F(q_i,t_{m-1})}{2},
\nonumber\\
\ee
we then have
\be
&&q^4\int_0^{t/2} ds (t-s)F(q,s)F^{(0)}(q,t-s)
\nonumber\\
&&=q^4 \Delta_t F^{(0)}(q,t)
\bigg[t\sum_{m=1}^{j/2}d\bar{F}(q,t_m) - \sum_{m=1}^{j/2}d\tilde{F}(q,t_m)\bigg].
\ee
The second integral similarly becomes
\be
\int_0^{t-t/2} ds (1+q^2s)F^{(0)}(q,s)\frac{\partial}{\partial s}F(q,t-s)
&\approx&\sum_{m=1}^{j-j/2}\bigg[\frac{F(q,t-t_m)-F(q,t-t_{m-1})}{q^2\Delta_t}\bigg]
\nonumber\\
&&\times [(2+q^2t_{m-1})e^{-q^2t_{m-1}}-(2+q^2t_m)e^{-q^2t_m}].
\nonumber\\
\ee

Plugging these terms back into Eq.~(\ref{eq:Ftilde}), we compactly write
\be
\tilde{F}[i][j] = E[i]F[i][j]+Z[i][j]
\ee
where
\be
E[i] = 1+F^{(0)}[i][[1]
\ee
and
\be
Z[i][j] &=& F^{(0)}[i][j]\bigg[1+(i\Delta_q)^2t+(i\Delta_q)^4\Delta_t
\sum_{m=1}^{j/2}\bigg(j\Delta_td\bar{F}[i][m]-d\tilde{F}[i][m]\bigg)\bigg]
\nonumber\\
&&-F^{(0)}[i][j-j/2]F[i][j/2][1+(i\Delta_q)^2(j-j/2)\Delta_q]-F[i][j-1]F^{(0)}[i][1]
\nonumber\\
&&+\bigg(\frac{1}{(i\Delta_q)^2\Delta_t}\bigg)\sum_{m=2}^{j-j/2}\bigg[F[i][j-m]-F[i][j-m-1]\bigg]
\nonumber\\
&&\times\bigg[(2+(i\Delta_q)^2(m-1)\Delta_t)e^{-(i\Delta_q)^2(m-1)\Delta_t}
-(2+(i\Delta_q)^2j\Delta_t)e^{-(i\Delta_q)^2m\Delta_t}\bigg]
\ee

At $q=0$, these reduce to $E[0] = 2$ and $Z[0][j] = -1$ yielding $\tilde{F}[0][j] = 1$
as expected.

\subsection{Updating the arrays at step doubling}
Just as the intermediate structure factor and memory function are updated as the step size is doubled, the dressed propagator arrays will also need to be updated. This is accomplished as follows:
\be
\bar{F}[i][j] &=& \bar{F}[i][2j]\\
\tilde{F}[i][j] &=& \tilde{F}[i][2j]\\
d\bar{F}[i][j] &=& \bigg(e^{-(i\Delta_q)^2(2j)\Delta_t}F[i][2j]
+ 4e^{-(i\Delta_q)^2(2j-1)\Delta_t}F[i][2j - 1]
\nonumber\\
&&+ e^{-(i\Delta_q)^2(2j-2)\Delta_t}F[i][2j - 2]\bigg)/6\\
d\tilde{F}[i][j] &=& \bigg((2j)\Delta_t e^{-(i\Delta_q)^2(2j)\Delta_t}F[i][2j]
+ 4(2j-1)\Delta_t e^{-(i\Delta_q)^2(2j-1)\Delta_t}F[i][2j - 1]
\nonumber\\
&&+ (2j-2)\Delta_t e^{-(i\Delta_q)^2(2j-2)\Delta_t}F[i][2j - 2]\bigg)/6.
\ee

\bibliography{references}
\end{document}